\documentclass[11pt, nofootinbib, showkeys
] {revtex4}

\pdfoutput=1
\usepackage{epsf,epsfig,subfigure,graphicx,amsmath,amssymb}

\usepackage{hyperref}

\usepackage{color}
\newcommand{\blue}{\textcolor{black}}
\newcommand{\red}{\textcolor{black}}
\newcommand{\magenta}[1]{{\color{black}#1}}
\newcommand{\pink}{\textcolor{black}}

\newcommand{\rred}{\textcolor{black}}
\definecolor{purple}{rgb}{0.87,0,1}
\newcommand{\purple}{\textcolor{black}}
\newcommand{\bblue}{\textcolor{black}}

\def\lsim{\mathrel{\rlap{\lower4pt\hbox{\hskip1pt$\sim$}}
    \raise1pt\hbox{$<$}}}         %less than or approx. symbol
\def\gsim{\mathrel{\rlap{\lower4pt\hbox{\hskip1pt$\sim$}}
    \raise1pt\hbox{$>$}}}         %greater than or approx. symbol

\newcommand{\df}{\text{d}}

\newcommand{\GeV}{\ensuremath{\,\text{GeV} }}
\newcommand{\nn}{\nonumber\\}

\newcommand{\al}[1]{\begin{align}#1\end{align}}

\newcommand{\paren}[1]{\left(#1\right)}
\newcommand{\fn}[1]{\!\left(#1\right)}
\newcommand{\sqbr}[1]{\left[#1\right]}
\newcommand{\br}[1]{\left\{#1\right\}}

\newcommand{\MSbar}{\ensuremath{\overline{\text{MS}}} }
\newcommand{\R}{\mathcal{R}}

\begin{document}

\vbox{
\hfill \hbox{\normalsize KUNS-2510, OU-HET/824-2014}
} \bigskip
\title{\bf \Large Higgs inflation from Standard Model criticality}

\author{
Yuta~Hamada$^{\dagger}$,
Hikaru~Kawai$^{\dagger}$, 
%Jinsu~Kim\thanks{E-mail: \tt kimjinsu@skku.edu},\\
Kin-ya~Oda$^\ddagger$, and
Seong~Chan~Park$^{\S}$}
\email{hamada@gauge.scphys.kyoto-u.ac.jp\\ hkawai@gauge.scphys.kyoto-u.ac.jp\\odakin@phys.sci.osaka-u.ac.jp\\ s.park@skku.edu}
%\bigskip
\address{$^{\dagger}$ Department of Physics, Kyoto University, Kyoto 606-8502, Japan\\
%\it %\normalsize
$^\ddagger$ Department of Physics, Osaka University, Osaka 560-0043, Japan\\
%\it %\normalsize
$^{\S}$ Department of Physics, Sungkyunkwan University, Suwon 440-746, Korea\\
and Korea Institute for Advanced Study, Seoul 130-722, Korea}

\vspace{1.0cm}
\begin{abstract}
\begin{center}
{\bf Abstract} 
\end{center}
\noindent \normalsize
The observed Higgs mass $M_H=125.9\pm0.4\GeV$ leads to the criticality of the Standard Model, that is, the Higgs potential becomes flat around the scale $10^{17\text{--}18}\GeV$ for the top mass $171.3$\,GeV. Earlier we have proposed a Higgs inflation scenario in which this criticality plays a crucial role. In this paper, we investigate \purple{the} detailed cosmological predictions of this scenario in light of the latest Planck and BICEP2 results.
%We find that this scenario can be consistent with the constraint from the running index too.
\purple{We also consider the Higgs portal scalar dark matter model, and compute the Higgs one-loop effective potential with the two-loop renormalization group improvement. We}
%We also compute the Higgs one-loop effective potential \purple{, including the Higgs portal scalar dark matter,}
find a constraint on the coupling between Higgs and dark matter \purple{which depends} on the inflationary parameters.
%In the Higgs portal scalar dark matter model, we have also obtained the high scale Higgs potential in terms of the coupling between the Higgs and the dark matter at the electroweak scale, using the two-loop renormalization group equations and the one-loop effective potential.
%\magenta{The constraint is given in terms of the effective Higgs quartic coupling at high scales.}
%\magenta{We also show its relation, in the Higgs portal scalar dark matter model, to the coupling between the dark matter and Higgs using the one-loop effective potential.}
%We also show the results including a Higgs portal scalar dark matter.
%We also present the predictions of the Higgs inflation in Higgs portal scalar Dark Matter model.
\end{abstract}

 %\pacs{95.35.+d, 95.85.Ry, 12.60.Cn,12.90.+b}

 %95.35.+d DM
 %98.62.Gq galactic halos
 %98.70.Rz gamma ray sources
 %95.85.Ry neutrino muon and other elem. particles,
 %11.30.Ly: other internal and higher symmetries
 %12.60.Cn: extension of EW sector
 %12.90.+b: Miscellaneous models
 %14.70.Pw: Other gauge bosons

 %\keywords{neutrino, dark matter, IceCube, WIMPZILLA, PeV events}

\maketitle
\newpage

\section{Introduction}
The observed value of the Higgs mass~\cite{PDG2014}\footnote{\red{The latest  values of the Higgs mass are $125.03^{+0.26}_{-0.27}\text{(stat)}^{+0.13}_{-0.15}\text{(syst)}$\,GeV (CMS)\cite{CMS2014} and  $125.36\pm 0.37 \text{(stat)}\pm 0.18\text{(syst)}$\,GeV (ATLAS)\cite{ATLAS2014}, which are consistent with each other and also with the PDG value we are using here.}}
\al{
M_H&=	125.9\pm0.4\GeV
		\label{current Higgs mass}
}
indicates that the Standard Model (SM) Higgs potential becomes small and flat at the scale around $10^{17\text{--}18}\,\text{GeV}$ \magenta{for the top mass 171.3\,GeV}; see e.g.~\cite{Holthausen:2011aa,Bezrukov:2012sa,Degrassi:2012ry,Alekhin:2012py,Masina:2012tz,Hamada:2012bp,Jegerlehner:2013cta,Buttazzo:2013uya,Branchina:2013jra,Kobakhidze:2014xda,Spencer-Smith:2014woa,Branchina:2014usa,Branchina:2014rva} for latest analyses.\footnote{
It is an intriguing fact that the bare Higgs mass also becomes small at the same scale~\cite{Hamada:2012bp,Hamada:2013cta,Masina:2013wja,Frandsen:2013bfa}; see also Refs.~\cite{Alsarhi:1991ji,Jones:2013aua,Haba:2013lga,Antipin:2013exa,Antipin:2013bya}. The running Higgs mass after the subtraction of the quadratic divergence is considered e.g.\ in Ref.~\cite{Bian:2013xra}; see also Refs.~\cite{Iso:2009ss,Iso:2009nw,Aoki:2012xs,Iso:2012jn,Hashimoto:2013hta,Hashimoto:2014ela,Chankowski:2014fva,Kobakhidze:2014afa}.
}
This fact suggests~\cite{Hamada:2013mya} that \red{the Higgs field \magenta{beyond the ultraviolet (UV) cutoff of the SM\purple{,} at the criticality~\cite{Froggatt:1995rt}\purple{,}} \red{may play the role of the slowly rolling} inflaton in the early universe; \magenta{see Ref.~\cite{Bezrukov:2007ep} for the original proposal to use the Higgs field for the cosmological inflation and also Refs.~\cite{Masina:2011aa,Masina:2011un,Masina:2012yd,Fairbairn:2014nxa} for the idea to use the false vacuum of the SM at criticality.} \magenta{Especially,} under the presence of the large non-minimal coupling $\xi\sim10^4$ between the Higgs field and the Ricci curvature, \magenta{there arises a 
%effective UV cutoff of the SM 
plateau in the SM effective potential
above the field value $\varphi\sim M_P/\sqrt{\xi}$}, and 
enough number of $e$-foldings is \magenta{achieved} without introducing any other field beyond the SM~\cite{Bezrukov:2007ep,Barvinsky:2008ia,DeSimone:2008ei,Bezrukov:2008ej,Bezrukov:2009db,Bezrukov:2010jz,Salvio:2013rja}.}
\pink{
In Ref.~\cite{Allison:2013uaa}, 
%an effort has been made to realize as small value of $\xi$ as possible by making the value of the effective quartic coupling $\lambda_\text{eff}$ very small at the inflationary scale; 
it has been shown by numerical analysis that smaller values of $\xi\sim 400$ and 90 are possible in the prescriptions I and II, respectively; see Sec.~\ref{inflation model} for what ``prescription'' means.
}

\magenta{In Ref.~\cite{Hamada:2014iga}, we have proposed to push the idea of Ref.~\cite{Hamada:2013mya} to use the criticality of the SM for the Higgs inflation scenario in order to accommodate a lower value of $\xi=7$--100, as well as a wider range of the tensor-to-scalar ratio $r\lesssim0.2$; see also Refs.~\cite{Cook:2014dga,Bezrukov:2014bra}.\footnote{\pink{
See Sec.~\ref{constraint on mu_min} for an explanation for the apparent discrepancy between the results from Refs.~\cite{Hamada:2014iga,Bezrukov:2014bra} and those from Ref.~\cite{Allison:2013uaa,Cook:2014dga}.
}}
Similar attempts \purple{have} been done in some extensions of the SM~\cite{Kawai:2014doa,Haba:2014zda,Hamada:2014iga,Ko:2014eia,Haba:2014zja,He:2014ora}.
There have also been different directions of the extension of the Higgs inflation involving higher dimensional operators~\cite{Germani:2010gm,Kamada:2010qe,Kamada:2012se,Chakravarty:2013eqa,Kamada:2013bia,Nakayama:2014koa,Lee:2014spa,Rubio:2014wta}.
See also Refs.~\cite{Hosotani:1985at,BenDayan:2009kv,Kubota:2011re,Ben-Dayan:2013eza,Jegerlehner:2014mua,Lindley:2014gia,Cheng:2014pna,Burgess:2014lza,Gong:2014cqa,Okada:2014lxa,Oda:2014rpa,Enqvist:2014tta,Enqvist:2014bua,Bamba:2014mua,Channuie:2014kda,Higaki:2014dwa,Kannike:2014mia,Gao:2014pca,Haba:2014sia,Rinaldi:2014gua,Chen:2014zoa,Bamba:2014daa,Postma:2014vaa,Xianyu:2014eba,Li:2014zfa,Inagaki:2014wva,Elizalde:2014xva,Ben-Dayan:2014iya,Bhattacharya:2014gva}.
}

\magenta{In this paper, we give detailed analyses of the Higgs inflation scenario proposed in Ref.~\cite{Hamada:2014iga} that utilizes the saddle point, at which both the first and second derivatives of the potential become very small.}
{The scale dependence of the effective quartic coupling $\lambda_\text{eff}$ is very important \purple{to determine the effective potential, whose behavior} around the saddle point is characterized by the minimum value $\lambda_\text{min}$ of the effective coupling $\lambda_\text{eff}$, the corresponding scale $\mu_\text{min}$, and the second derivative $\beta_2$ of $\lambda_\text{eff}$ around $\mu_\text{min}$, in addition to $\xi$.}
We examine the predictions of this model on spectral index $n_s$, tensor to scalar ratio $r$, and the running of spectral index \purple{$\df n_s/\df\ln k$}.

\purple{We also estimate how small the higher dimensional Planck-suppressed operators must be in order to maintain the observed values of the cosmic microwave background (CMB).}
%\magenta{We can further explore the modifications of the Higgs potential from Planck scale physics through its effects on the cosmic microwave background (CMB).}
%In this paper, we %introduce 
%\purple{estimate} this effect by %\magenta{minimally} 
%taking into account the Planck-suppressed 
\purple{For that purpose, we pick up} \magenta{the six-dimensional operator $\varphi^6/M_P^2$ in the Jordan-frame potential} as a \purple{concrete} example\purple{, and}
\magenta{compute the CMB spectral indices.

We also evaluate the relation \purple{between the high-scale parameters $\mu_\text{min}$, $\beta_2$ and} the low energy parameters in the SM\purple{, as well as} in the Higgs portal scalar dark matter (DM) model, using one-loop effective potential and the two-loop renormalization group equation (RGE).
}

This paper is organized as follows.
In Section 2, we review \magenta{
the criticality, namely the flatness and smallness, of the SM Higgs potential around the scale $10^{17\text{--}18}\GeV$.
In Section 3, we review the Higgs inflation scenario in a wider perspective.
%We explain the Higgs inflation model and review the original Higgs inflation~\cite{Bezrukov:2007ep} in Section 3.
In Section 4, we investigate the predictions of this model in detail.
In Section 5, we consider the extension with the Higgs portal scalar DM.}
We summarize our results in the last section.

%%%%%%%%%%%%%%%%%%%%%%%%%%%%%%%%%%%%%%%%
\section{Standard Model Higgs potential}\label{SM}
In the SM on the flat spacetime background, the one-loop effective potential calculated in the \MSbar scheme in the Landau gauge is
\al{
V%_\text{eff}
	&=	V_\text{tree}+\Delta V_\text{1-loop},
		\label{one-loop minimal}
}
with
\al{\label{tree lambda}
V_\text{tree}
&=
\blue{
e^{4\Gamma(\varphi)}
}
{\lambda(\mu)\over 4}\varphi^4,\\
\Delta V_\text{1-loop}
	&=	\blue{
e^{4\Gamma(\varphi)}
}
\bigg\{
-\frac{3m_t(\varphi)^4}{16\pi^2}\left(\ln\frac{m_t(\varphi)^2}{\mu^2}-\frac{3}{2}
+
\blue{
2\Gamma(\varphi)
}
\right)
		\nn
	&\quad
		+\frac{6m_W(\varphi)^4}{64\pi^2}\left(\ln\frac{m_W(\varphi)^2}{\mu^2}-\frac{5}{6}
+
\blue{
2\Gamma(\varphi)
}
\right)
		+\frac{3m_Z(\varphi)^4}{64\pi^2}\left(\ln\frac{m_Z(\varphi)^2}{\mu^2}-\frac{5}{6}
+
\blue{
2\Gamma(\varphi)
}
\right)
\bigg\},
		\label{1-loop lambda}
\\
\Gamma(\varphi)
&=\int^{\varphi}_{M_t}\gamma\ d\ln \mu,
\label{wavefunction}
\\
\gamma&=\frac{1}{(4\pi)^2}\left(
\frac{9}{4}g_2^2+{3 \over 4}g_Y^2-3y_t^2
\right),
}
where $m_W(\varphi)=g_2\varphi/2$, $m_Z(\varphi)=\sqrt{g_Y^2+g_2^2}\, \varphi/2$, and $m_t(\varphi)=y_t \varphi/\sqrt{2}$.
We have neglected the effects from the loops of the Higgs and would-be Nambu-Goldstone bosons since we are interested in the scale where $\lambda$ becomes small. \red{We also neglect the quadratic term; the bare Higgs mass is 
%screened 
%irrelevant
canceled by the loop effect
%not important 
at low energies; see e.g.\ Appendix B in Ref.~\cite{Hamada:2013mya}.}

We define the effective quartic coupling as~\cite{Degrassi:2012ry}
\al{
V(\varphi)
	&=	{\lambda_\text{eff}(\varphi,\mu)\over4}\varphi^4.
}
At the one-loop level,
\al{\label{one-loop log}
\lambda_\text{eff}(\varphi,\mu)
	&=\blue{
e^{4\Gamma(\varphi)}
}
	\lambda(\mu)+
\blue{
e^{4\Gamma(\varphi)}
}
{1\over16\pi^2}\Bigg[
		-3y_t^4 \paren{
			\ln{y_t^2\varphi^2\over2\mu^2}
			-{3\over2}\blue{+2\Gamma(\varphi)}
			}\nn
	&\phantom{=	\lambda(\mu)+{1\over16\pi^2}\Bigg[}
		+{3g_2^4\over 8}\paren{
			\ln{g_2^2\varphi^2\over4\mu^2}
			-{5\over6}\blue{+2\Gamma(\varphi)}
			}
		+{3\paren{g_Y^2+g_2^2}^2\over 16}\paren{
			\ln{\paren{g_Y^2+g_2^2}\varphi^2\over4\mu^2}
			-{5\over6}\blue{+2\Gamma(\varphi)}
			}
		\Bigg],
}
where we have made the scale dependence explicit in the right hand side while omitting it in $y_t$, $g_2$, and $g_Y$, which corresponds to the two-loop corrections.
%We also neglect the field renormalization of $\varphi$ since it gives a one-loop correction only in $V_\text{tree}$, which is itself tuned to be of the order of one-loop potential in our scenario.\blue{delete?}

In the SM on the flat spacetime background, $\Delta V_\text{1-loop}$ is minimized by
\al{
\ln{\varphi^2\over\mu^2}
	&=	{C_t^2\paren{-\ln C_t+{3\over2}-\blue{2\Gamma}}-2C_W^2\paren{-\ln C_W+{5\over6}-\blue{2\Gamma}}
			-C_Z^2\paren{-\ln C_Z+{5\over6}-\blue{2\Gamma}}
			\over C_t^2-2C_W^2-C_Z^2}\nn
	&=	{\paren{-\ln C_t+{3\over2}-\blue{2\Gamma}}-2{C_W^2\over C_t^2}\paren{-\ln C_W+{5\over6}-\blue{2\Gamma}}
			-{C_Z^2\over C_t^2}\paren{-\ln C_Z+{5\over6}-\blue{2\Gamma}}
			\over 1-2{C_W^2\over C_t^2}-{C_Z^2\over C_t^2}},
			\label{renormalization condition}
}
where $C_W=g_2^2/4$, $C_Z=\paren{g_Y^2+g_2^2}/4$, and $C_t=y_t^2/2$.
Around $\mu_\text{min}\sim 10^{17\text{--}18}$GeV, %we have %$y_t\simeq0.40$, $g_Y\simeq0.45$, $g_2\simeq0.52$ \blue{and $\Gamma\simeq$} and
Eq.~\eqref{renormalization condition} leads to %$\ln(\varphi^2/\mu^2)\simeq2.9$, namely 
$\mu\simeq0.23\varphi$. %\footnote{
%To rewrite:
%\als{
%\ln{C_t\varphi^2\over\mu^2}
%	&=	{\paren{-2{C_W^2\over C_t^2}-{C_Z^2\over C_t^2}}\ln C_t
%		+{3\over2}
%		-2{C_W^2\over C_t^2}\paren{-\ln C_W+{5\over6}}
%			-{C_Z^2\over C_t^2}\paren{-\ln C_Z+{5\over6}}
%			\over 1-2{C_W^2\over C_t^2}-{C_Z^2\over C_t^2}}.
%}
%An exponential of minus of the right hand side gives what is called in Ref.~\cite{Bezrukov:2014bra} $\alpha\simeq0.7$.
%}
However, because the difference of the numerical values of the one-loop effective potential for  $\mu=\varphi$ and $\mu=0.23\varphi$ is
%between one-loop effective potential with $\mu=\varphi$ and one with %$\mu=0.24\varphi$ 
negligibly small, we use $\mu=\varphi$ hereafter in this section.

Then, we obtain 
\al{
V(\varphi)
	&=	{\lambda_\text{eff}(\mu=\varphi)\over4}\varphi^4,
}
where $\lambda_\text{eff}(\mu)$ is written by
\al{\label{effective coupling}
\lambda_\text{eff}(\mu)&=
\blue{
e^{4\Gamma}
}
\lambda(\mu)
%-{3\over 16\pi^2}y_t(\mu)^4 \left(-{3\over2}+\ln{y_t(\mu)^2\over2}\right),
+\blue{
e^{4\Gamma}
}
{1\over16\pi^2}\Bigg[
		-3y_t^4 \paren{
			\ln{y_t^2\over2}
			-{3\over2}
                     \blue{+2\Gamma}
			}\nn
	&\phantom{=	\lambda(\mu)+{1\over16\pi^2}\Bigg[}
		+{3g_2^4\over 8}\paren{
			\ln{g_2^2\over4}
			-{5\over6}
                     \blue{+2\Gamma}
			}
		+{3\paren{g_Y^2+g_2^2}^2\over 16}\paren{
			\ln{\paren{g_Y^2+g_2^2}\over4}
			-{5\over6}
                     \blue{+2\Gamma}
			}
		\Bigg],
}
at the one-loop level.

\rred{The effective coupling $\lambda_{\rm eff}$ is quartically sensitive to $y_t$ thus the  top quark mass, $M_t$, which is scheme dependently defined.  The actual value of $M_t$ is known with large uncertainties at the level of GeV scales depending on the measurements:
\begin{align}
M_t^\text{pole}
	&= 
\begin{cases}
 171.2\pm2.4\GeV, & \text{MITP \cite{Moch:2014tta}},\\
 176.7^{+4.0}_{-3.4}\GeV, & \text{PDG \cite{Agashe:2014kda}}, \\ 
\end{cases} \label{pole mass}\\
M_t^\text{Pythia}
	&=	
\begin{cases}
 173.21\pm 0.51 \pm 0.71 \GeV, & \text{direct measurement, PDG \cite{Agashe:2014kda}}, \\ 
 174.98 \pm0.76 \GeV, & \text{D0 \cite{Dzerotop}},\\
 174.34\pm0.64\GeV, &\text{D0+CDF \cite{Tevatron:2014cka}},\\
 173.34 \pm0.76\GeV, & \text{ATLAS \cite{ATLAS:2014wva}},\\ 
 172.38 \pm 0.10 \pm 0.65 \GeV, & \text{CMS \cite{CMStop}}.\\
\end{cases}\label{Pythia mass}
\end{align}
One should note that the ``directly measured value'' in Eq.~\eqref{Pythia mass} obtained by Tevatron (D0 and CDF) and by LHC (ATLAS and CMS) is indeed a parameter in Monte Carlo simulation code~\cite{Alekhin:2012py,Kawabataa:2014osa}, so-called the Pythia mass~\cite{Kawabata:Talk}, whose physical relation to the pole and $\overline{\text{MS}}$ masses is not well established. In discussing the Higgs inflation near criticality, however, the only important fact is that the critical value for the pole mass $M_t\simeq171.3\GeV$, shown just below, is perfectly consistent with both the MITP and PDG within $2\sigma$ confidence level. In below, we will take %consider $2\sigma$ band with 
the MITP value as a benchmark.}

%We plot tree level $\lambda_\text{eff}$, one-loop level $\lambda_\text{eff}$, tree and one-loop level $10\,\df \lambda_\text{eff}/\df\ln\mu$ in Fig.~\ref{lambda}. 
%The band corresponds $95\%$ CL deviation of top-quark pole mass, where
%\al{
%M_t=173.3\pm2.8\GeV,
%\magenta{M_t=171.2\pm2.4\GeV,}
%	\label{current top mass}
%}
%at the $1\sigma$ level~\magenta{\cite{Moch:2014tta}};
%\pink{this is perfectly consistent to the result with the bigger error $176.7^{+4.0}_{-3.4}\GeV$ quoted in Ref.~\cite{Agashe:2014kda}.}
%We note that the ``top mass'' that is precisely determined to be $173.34\pm0.76\GeV$~\cite{ATLAS:2014wva} \magenta{and $174.34\pm0.64\GeV$~\cite{Tevatron:2014cka}} \pink{(and is quoted in Ref.~\cite{Agashe:2014kda} as $173.21\pm0.51\pm0.71\GeV$)} is mere a parameter in a Monte Carlo simulation code~\cite{Alekhin:2012py,Kawabataa:2014osa}, so-called the Pythia mass~\cite{Kawabata:Talk}, whose relation to the pole mass is not clear.\footnote{
%\red{The latest top-quark \pink{Pythia} mass is measured \pink{with rather scattered values:}
%to be within a rather large range:
%\pink{$172.38 \pm 0.10\,\text{(stat.)}\pm 0.65\,\text{(syst.)}\GeV$} by CMS~\cite{CMStop} and $174.98\pm0.76$\,GeV by D0~\cite{Dzerotop}.}}
%\pink{Here the only important fact is that the critical value for the pole mass $M_t\simeq171.3\GeV$ that we see below is within the given error band at the confidence level you want, say 2$\sigma$.}

\begin{figure}[tn]
\begin{center}
\hfill
\includegraphics[width=.6\textwidth]{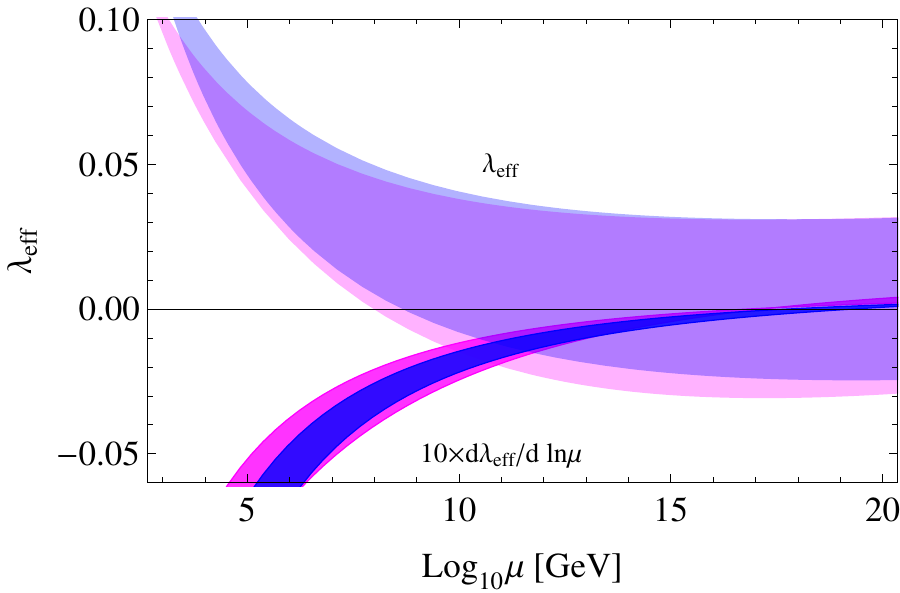}
%\hfill
%\includegraphics[width=.32\textwidth]{looppotential.pdf}
%\hfill
%\includegraphics[width=.32\textwidth]{mH_vs_mumin.pdf}
%\hfill
%\includegraphics[width=.49\textwidth]{mH_vs_mumin2.pdf}
\hfill\mbox{}\\
\caption{The light red (lower) and blue (upper) bands are 2-loop RGE running of $\lambda(\mu)$ and $\lambda_\text{eff}(\mu)$~\eqref{effective coupling} 
%from the tree level potential~\eqref{tree lambda} and from the 1-loop effective potential~\eqref{tree lambda} and \eqref{1-loop lambda}
, respectively.
The dark red (upper) and blue (lower) bands are the beta function times ten $10\times\df\lambda_\text{eff}/\df\ln\mu$ evaluated at the tree and 1-loop levels, respectively.
We take $M_H=125.9\GeV$ and $\alpha_s=0.1185$. The band corresponds to $95\%$ CL deviation of $M_t$~\cite{Moch:2014tta}; see Eq.~\eqref{pole mass}.
%Note that the graviton loop effects are neglected in the plot, and hence the region $\varphi\gtrsim 1/\sqrt{32\pi G}=1.2\times10^{18}\GeV$ cannot be trusted; this part has been included in the plot to make the fact that there is minimum around $10^{17}\GeV$ regardless of .
}
\label{lambda}
\end{center}
\end{figure}

\begin{figure}[tn]
\begin{center}
\hfill
\includegraphics[width=.49\textwidth]{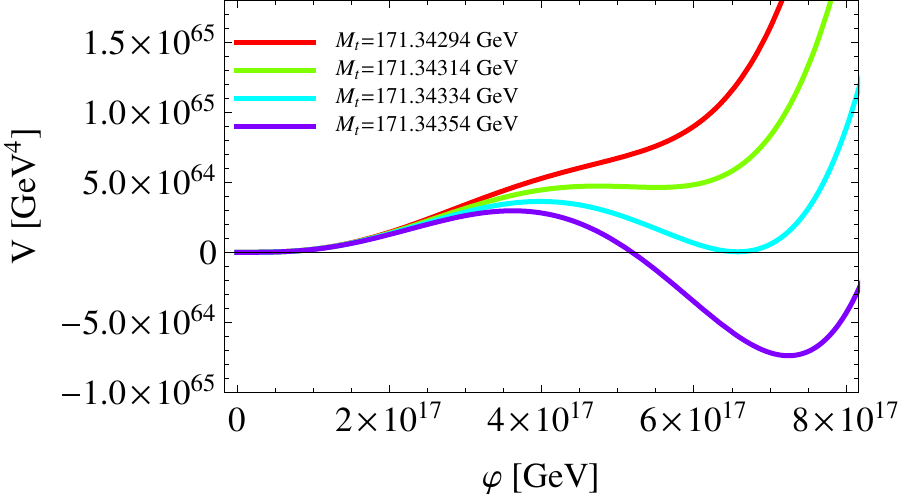}
\hfill
\includegraphics[width=.49\textwidth]{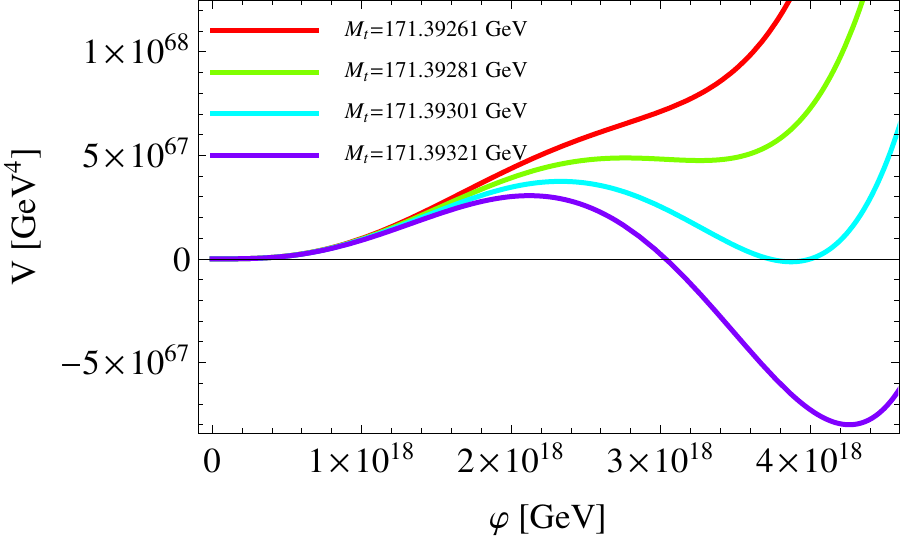}%\\
%\hfill
%\includegraphics[width=.49\textwidth]{looppotential2.pdf}
%\hfill
%\includegraphics[width=.49\textwidth]{mH_vs_mumin2.pdf}
\hfill\mbox{}\\
\caption{Left: The tree level Higgs potential~\eqref{tree lambda} as a function of Higgs field $\varphi$. Right: The one-loop Higgs potential~\eqref{tree lambda} and \eqref{1-loop lambda}. We take $M_H=125.9\GeV$ and $\alpha_s=0.1185$.}
\label{SMpotential}
\end{center}
\end{figure}

In Fig.~\ref{lambda}, we can see that $\lambda_\text{eff}$ has the minimum around $10^{17\text{--}18}\GeV$.
Interestingly, \purple{if $M_t\simeq171.3\GeV$, the \purple{minimum value} of $\lambda_\text{eff}$ \purple{becomes} zero around the scale $10^{17\text{--}18}\GeV$, and}
%\red{,which is consistent with the latest CMS value}.
the Higgs potential \blue{has a plateau} around $10^{17\text{--}18}\GeV$ as shown in Fig.~\ref{SMpotential}.\footnote{\label{gauge dependence}
\magenta{
%In Fig.~\ref{SMpotential}, we see that the field value $\varphi$ that realizes the plateau of the potential shifts by an order of magnitude from the tree to 1-loop potentials.
It has been known that such a position of plateau is unphysical and can vary by an order of magnitude depending on the gauge choice~\cite{DiLuzio:2014bua}.
The gauge dependence of the effective potential can be absorbed by a field redefinition~\cite{Nielsen:2014spa}.
The eventual field equation for $\varphi$ should not depend on such a choice, but the field value here necessarily contains this amount of uncertainty.
\pink{See also Refs.~\cite{Andreassen:2014eha,Andreassen:2014gha} for a further account on the gauge (in)dependence.}
}
}

%We can expand the running coupling $\lambda(\mu)$ around its minimum~\cite{Hamada:2014xka}
%\al{
%\lambda(\mu)&\simeq\lambda_\text{min}+\frac{\beta_2}{\paren{16\pi^2}^2}\left(\ln\frac{\mu}{\mu_\text{min}}\right)^2,
%}
%where $\beta_2$ is an $O(1)$ quantity and $\lambda_\text{min}$ and $\mu_\text{min}$ are functions of the top and Higgs masses. 
%In the SM, $\beta_2\simeq$, see e.g.\ Appendix A in Ref.~\cite{Hamada:2013mya}, and $\mu_\text{min}$ is $10^{17\text{--}18}$GeV~\cite{Holthausen:2011aa,Bezrukov:2012sa,Degrassi:2012ry,Alekhin:2012py,Masina:2012tz,Hamada:2012bp,Jegerlehner:2013cta,Buttazzo:2013uya}. 

Let us expand the effective potential of the Higgs field $V_\text{eff}(\varphi)$ on the flat space-time background around its minimum in terms of $\ln\varphi$:
\al{\label{expansion}
V(\varphi)
	&=	{\lambda_\text{eff}(\mu=\varphi)\over4}\varphi^4,	&
\lambda_\text{eff}(\mu)
	&=	\lambda_\text{min}+\sum_{n=2}^\infty{\beta_n\over\paren{16\pi^2}^n}\paren{\ln{\mu\over\mu_\text{min}}}^2,
}
where the overall factor $\varphi^4$ is put to make the expansion \purple{well-behaved}.
In the potential analysis around the minimum, we can safely neglect the higher order terms with $n\geq3$, and will omit them hereafter.
By tuning the top mass for a given Higgs mass, we can obtain arbitrarily small $\lambda_\text{min}$.
This fact is crucial for our inflation scenario.

\begin{figure}[tn]
\begin{center}
%\hfill
%\includegraphics[width=.32\textwidth]{mH_vs_Mt.pdf}
%\hfill
%\includegraphics[width=.32\textwidth]{mH_vs_beta2.pdf}
%\hfill
%\includegraphics[width=.32\textwidth]{mH_vs_mumin.pdf}\\
%\hfill
%\includegraphics[width=.49\textwidth]{mH_vs_mumin2.pdf}
\hfill
\includegraphics[width=.32\textwidth]{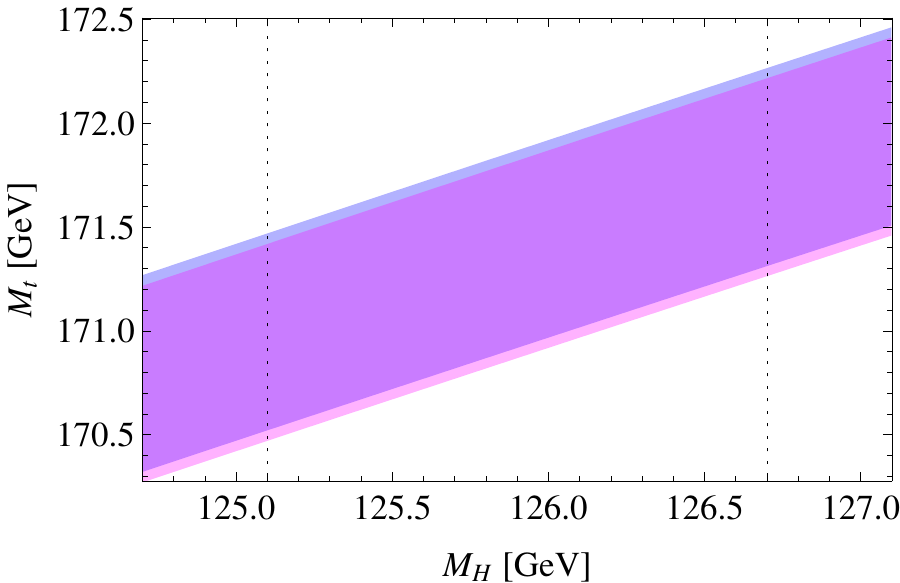}
\hfill
\includegraphics[width=.32\textwidth]{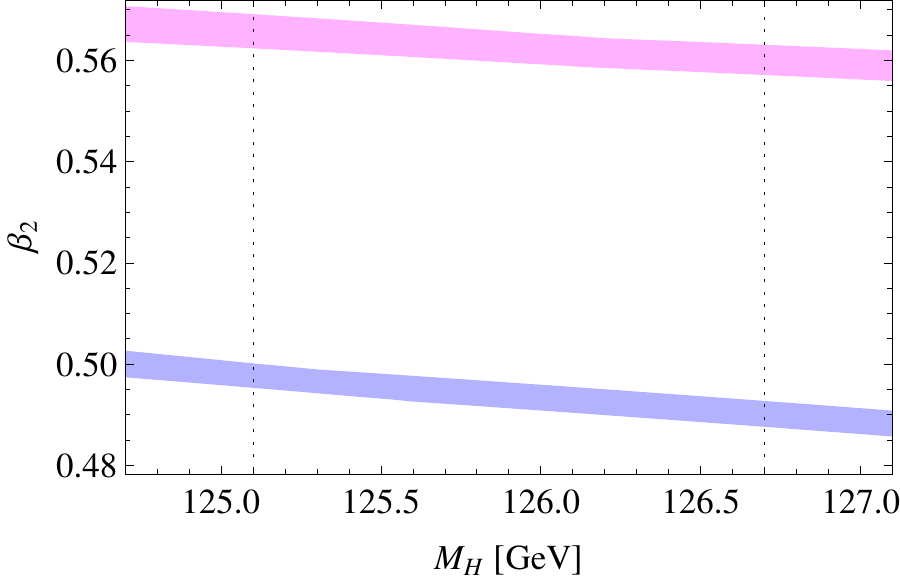}
\hfill
\includegraphics[width=.32\textwidth]{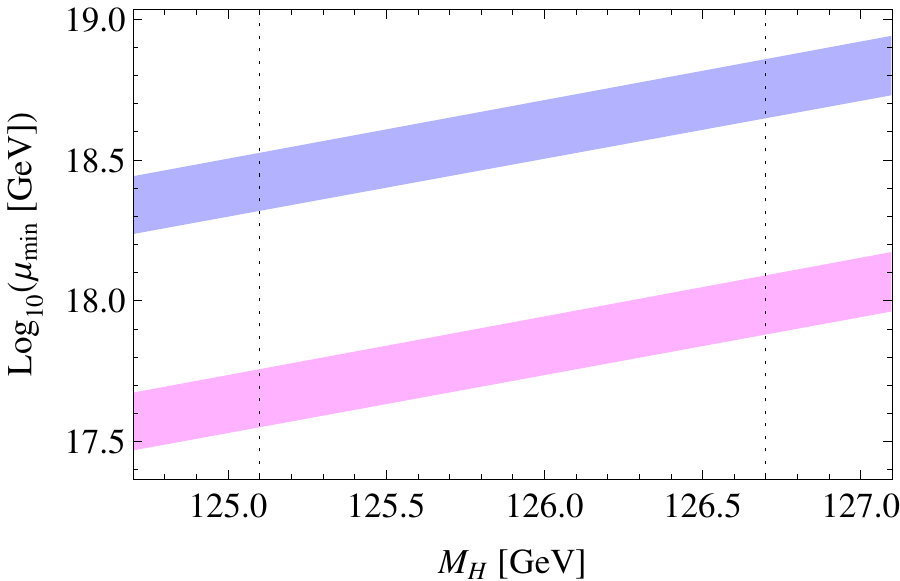}
\hfill\mbox{}\\
\caption{
$M_t$ (left), $\beta_2$ (center), and $\mu_\text{min}$ (right) that realize the condition $\lambda_\text{min}=\lambda_c$ are plotted as functions of $M_H$.
We have imposed the condition $\lambda_\text{min}=\lambda_c$ using the tree-level potential~\eqref{tree lambda} and the one-loop one, \eqref{tree lambda} and \eqref{1-loop lambda}, for the red and blue bands, respectively.
(The one-loop blue band is the upper one for left and right, whereas the lower for center.)
The width of the bands corresponds to the 95\% CL of $\alpha_s(M_Z)$.
Dotted lines show the current 95\% CL for $M_H$; see Eq.~\eqref{current Higgs mass}.
}
\label{SMvalue}
\end{center}
\end{figure}

We note that \magenta{for the potential to be monotonically increasing~\cite{Hamada:2014iga}, $\lambda_\text{min}$ must be larger than a critical value~$\lambda_c$:}
\al{
\magenta{\lambda_\text{min}\geq \lambda_c:=\frac{\beta_2}{\paren{64\pi^2}^2}.} \label{lambda_c}
}
\magenta{When $\lambda_\text{min}$ saturates this inequality,}
\al{
\lambda_\text{min}=\lambda_c,
	\label{lambda at critical}
}
there appears a \magenta{true} saddle point of the potential \magenta{$V_\varphi=V_{\varphi\varphi}=0$}.
\magenta{We will see in Section~\ref{Prescription I} that in the prescription I, this value $\lambda_c$ also gives the true saddle point of the modified potential: $U_\varphi=U_{\varphi\varphi}=0$.}\footnote{Numerical difference between the results from the condition $\lambda_\text{min}=0$ and from Eq.~\eqref{lambda at critical} is much smaller than the deviation coming from the $\alpha_s(M_Z)$ error.
We have imposed $\lambda_\text{min}=0$ within a precision of $10^{-5}$ in the actual numerical computation in writing Fig.~\ref{SMvalue}. Note that $\lambda_c=2.5\times10^{-6}\beta_2$.
}

In the left, center, and right of Fig.~\ref{SMvalue}, we plot $M_t$, $\beta_2$, and $\mu_\text{min}$, respectively, \magenta{with the critical value of $\lambda_\text{min}$ given in Eq.~\eqref{lambda at critical}}.
The band corresponds to the 95\% CL for the strong coupling constant measured at $\mu=M_Z$, where
\al{
\alpha_s(M_Z)=0.1185\pm0.0006
}
at the 1$\sigma$ level~\cite{PDG2014}.
We see that $\beta_2$ does not depend much on $M_H$. %\footnote{
%In contrast, %the position of the minimum of $\lambda(\mu)$
%}
In the following figures except Fig.~\ref{scalar}, we take a reference value $\beta_2=0.5$.
\footnote{
We have checked that the changes of spectral index, its running, its running of running, and tensor-to-scalar ratio are hardly seeable when we vary $\beta_2$.}
$\mu_\text{min}$ changes by an order of magnitude when one includes the one-loop corrections to the effective potential as shown in the right of Fig.~\ref{SMvalue}. 
%$\mu_\text{min}$ does not change by the two-loop corrections
The two-loop corrections are negligible compared with the one-loop corrections; see e.g.\ Ref.~\cite{Degrassi:2012ry}.
%One might worry about further two-loop corrections, but it is known to be small}
In Fig.~\ref{SMvalue}, we see that $\beta_2$ and $\mu_\text{min}$ \purple{differ} between at tree and one-loop levels, but
note that $M_t$ is almost identical at both levels.

%%%%%%%%%%%%%%%%%%%%%%%%%%%%%%%%%%%%%%%%
\section{Inflation model}\label{inflation model}
Let us consider the effective action of the SM-gravity system in the local potential approximation.
As we are interested in the spatially constant field configuration and the case where the Hubble parameter is much smaller than the Planck scale, we restrict ourselves to the terms containing up to second derivative of the fields.
We can write down the effective action schematically as\footnote{
In the Letter~\cite{Hamada:2014iga}, we have used $h$ instead of $\varphi$.
}
\al{
S	&=	\int\df^4x\sqrt{-g}\Bigg[
			{M_P^2\over2}\,A(\varphi)\,\R\nn
	&\phantom{=	\int\df^4x\sqrt{-g}\Bigg[}
			-{1\over2}\,B(\varphi)\,g^{\mu\nu}
\blue{\paren{
\partial_\mu \varphi\partial_\nu \varphi
+A_\mu A_\nu \varphi^2}
}
%\paren{\partial_\mu \varphi-iA_\mu\varphi}
%\paren{\partial_\nu \varphi+iA_\nu\varphi}
			-V(\varphi)\nn
	&\phantom{=	\int\df^4x\sqrt{-g}\Bigg[}
			-C(\varphi)\,\overline{\psi}\gamma^\mu D_\mu\psi
			-{y\over\sqrt{2}}\,D(\varphi)\,\paren{\varphi\overline{\psi}\psi+\text{h.c.}}\nn
	&\phantom{=	\int\df^4x\sqrt{-g}\Bigg[}
			-{E(\varphi)\over4g_A^2}\,F_{\mu\nu}F^{\mu\nu}
			\Bigg],
			\label{Jordan frame action}
}
where $g_A$ and $y$ are gauge and Yukawa couplings, respectively, $M_P:=1/\sqrt{8\pi G}=2.4\times10^{18}\GeV$ is the reduced Planck scale, $\varphi$ is the physical (real) Higgs field, and
\al{
A(\varphi)
	&=	1+a_2{\varphi^2\over M_P^2}+a_4{\varphi^4\over M_P^4}+\cdots,	&
B(\varphi)
	&=	1+b_2{\varphi^2\over M_P^2}+b_4{\varphi^4\over M_P^4}+\cdots,	&
\text{etc.,}
		\label{coefficients}
}
with $a_2$, \dots, $b_2$, \dots, etc.\ being dimensionless constants.
Generically the potential $V$ also contains higher dimensional terms
\al{
V(\varphi)
	&=	{m^2\over2}\varphi^2+{\lambda\over4}\varphi^4
		+\paren{\lambda_6{\varphi^6\over M_P^2}+\lambda_8{\varphi^8\over M_P^4}+\cdots}.
		\label{higher order potential}
}

We can recast the Jordan frame action~\eqref{Jordan frame action} by the field redefinition
\al{
g^E_{\mu\nu}
	&=	A(\varphi)\,g_{\mu\nu},
}
to get \red{the action in the Einstein frame}
\al{
S	&=	\int\df^4x\sqrt{-g_E}\Bigg[
			{M_P^2\over2}\,\R_E\nn
	&\phantom{=	\int\df^4x\sqrt{-g}\Bigg[}
			-{1\over2}\sqbr{{B(\varphi)\over A(\varphi)}+{3\over2}{B(\varphi)\,A'(\varphi)^2\over A(\varphi)^2}}\,
				g_E^{\mu\nu}
\blue{
\partial_\mu \varphi \partial_\nu \varphi
-{1\over2}\sqbr{{B(\varphi)\over A(\varphi)}}g_E^{\mu\nu}
A_\mu A_\nu \varphi^2
}					
%\paren{\partial_\mu \varphi-iA_\mu\varphi}
%\paren{\partial_\nu \varphi+iA_\nu\varphi}
			-{V(\varphi)\over A(\varphi)^2}\nn
	&\phantom{=	\int\df^4x\sqrt{-g}\Bigg[}
			-{C(\varphi)\over A(\varphi)^{3/2}}\overline{\psi}\gamma_E^\mu D_\mu\psi
			-{y\over\sqrt{2}}{D(\varphi)\over A(\varphi)^2}\paren{\varphi\overline{\psi}\psi+\text{h.c.}}\nn
	&\phantom{=	\int\df^4x\sqrt{-g}\Bigg[}
			-{E(\varphi)\over4g_A^2}\,F_{\mu\nu}F_E^{\mu\nu}
			\Bigg],
			\label{Einstein frame action}
}
see e.g.\ Refs.~\cite{Shapiro:1995kt,Park:2008hz}.

By the field redefinition:
\al{
{\df\chi\over\df\varphi}
	&=	\sqrt{{B(\varphi)\over A(\varphi)}+{3\over2}{B(\varphi)A'(\varphi)^2\over A(\varphi)^2}},	&
\widetilde\psi
	&=	{C(\varphi)^{1/2}\over A(\varphi)^{3/4}}\psi,
}
we get the canonically normalized kinetic term for $\chi$ and $\widetilde\psi$.\footnote{
There appear extra derivative terms from the kinetic term of the fermion.
We neglect such terms, since we are interested in the expression of the fermion mass for a constant background field $\varphi$ and for the Hubble parameter much smaller than the Planck scale.
}
For a given background field $\varphi$ in the Jordan frame, the effective mass for the canonically normalized field $\widetilde\psi$ is
\al{
m_{\widetilde\psi}
	&=	{y\varphi\over\sqrt{2}}{D(\varphi)\over\sqrt{A(\varphi)}\,C(\varphi)}.
		\label{fermion mass}
}
Similarly, the effective mass for a canonically normalized gauge field is
\al{
m_{\widetilde A}
	&=	g_A\varphi\sqrt{{B(\varphi)\over A(\varphi)\,E(\varphi)}%+{3\over2}{B(\varphi)A'(\varphi)^2\over A(\varphi)^2\,E(\varphi)}
}.
		\label{gauge boson mass}
}
For later convenience, we define the Einstein frame potential
\al{\label{U potential}
U(\varphi)
	&:=	{V(\varphi)\over {A(\varphi)}^2}.
}

In the original version of the Higgs inflation~\cite{Bezrukov:2007ep,Bezrukov:2009db,Bezrukov:2010jz}, it is assumed that $\xi:=a_2$ happens to be large: $\xi\sim 10^4$, whereas the other couplings are not much larger than unity: $\xi\gg a_2,a_4,\dots; b_2,\dots$ etc.
In that limit, we can write
\al{
A(\varphi)
	&=	1+{\xi\varphi^2\over M_P^2},	&
B(\varphi)
	=	C(\varphi)=D(\varphi)=E(\varphi)
	&=	1,   &
	\lambda_6=\lambda_8=\dots
	&=	0.
		\label{minimal coefficients}
}
As a side remark, we note that we can instead assume $b_2\sim10^5$ while keeping all other coefficients, including $a_2$, not much larger than unity in order to realize another version of Higgs inflation~\cite{Nakayama:2014koa}. It may be interesting to look for more possibilities of putting a large number in other places.
In this paper, we restrict ourselves to more conventional set of the non-minimal couplings~\eqref{minimal coefficients}, and later take into into account the term $\lambda_6\varphi^6$ in the potential~\eqref{higher order potential} as a next step.

\bblue{
For $\varphi\gg M_P/\sqrt{\xi}$, we have $\df\chi/\df\varphi\simeq\sqrt{6}M_P/\varphi$, \purple{which leads to} $\varphi\simeq{M_P\over\sqrt{\xi}}\exp\paren{\chi/\sqrt{6}M_P}$, and the potential \purple{becomes}~\cite{Bezrukov:2007ep}
}
\al{
U(\chi)
	&=	{V\over \paren{1+e^{2\chi/\sqrt{6}M_P}}^2}.
}
The analysis of this model without taking into account the running of $\lambda$ gives following predictions~\cite{Bezrukov:2007ep}
\al{
n_s&=1-6\epsilon_V+2\eta_V\simeq0.967,\nn
r&=16\epsilon_V\simeq3\times10^{-3},\nn
\frac{\df n_s}{\df\ln k}
	&=	16\epsilon_V\eta_V-24\epsilon_V^2-2\zeta_V^2\simeq-5.4\times10^{-4},
}
where 
\al{
\epsilon_V
	&=	\frac{M_P^2}{2}\left(\frac{\df U/\df\chi}{U}\right)^2,
		\label{epsilon defined}\\
\eta_V
	&=	M_P^2\frac{\df^2U/\df\chi^2}{U},\\
\zeta_V^2
	&=	M_P^4 \frac{(\df^3U/\df\chi^3) (\df U/\df\chi)}{U^2}.
}
%In this paper we assume that $f_1$, $f_2$, \dots are of order unity.

\blue{As is seen in Eq.~\eqref{one-loop log},
%, for a large Higgs field value $\varphi^2\gg\ab{m^2}$, 
the loop corrections to the effective potential contain large logarithms.
%We denote them by 
They can be written as $\ln\fn{M(\varphi)/\mu}$,} where $\mu$ is the renormalization scale and $M(\varphi)$ stands for the field-dependent mass of the particle running in the loop, namely the top quark and the gauge bosons. %in the case of SM.
The problem is that there are two possibilities in defining the field dependent mass~\cite{Bezrukov:2009db}.
In the so-called prescription I, we use the field dependent mass in the Einstein frame, as in Eqs.~\eqref{fermion mass} and \eqref{gauge boson mass}, whereas in the prescription II, we use the ones in the Jordan frame, namely $m_\psi=y\varphi/\sqrt{2}$ and $m_A=g_A\varphi$.
%In Ref.~\cite{George:2013iia}, the authors claim that the apparent difference of the results is due to the backreaction of the gravity and that one should use the prescription~I if one performs a naive computation without taking into account the backreaction.
%The argument of Ref.~\cite{George:2013iia} seems plausible but not yet fully conclusive.
We leave the possibilities open for future research and present our results for both prescriptions.

\blue{As we have done below Eq.~\eqref{renormalization condition},} in either prescriptions I or II, we can drop the gauge and Yukawa couplings in the field dependent mass.
%, that is, 
For the prescription I\purple{,} assuming the minimal set of coefficients~\eqref{minimal coefficients}, we put
\al{
\mu	&=	\frac{\varphi}{\sqrt{1+\xi \varphi^2/M_P^2}}
		\label{prescription I}
}
and for the prescription II,
\al{
\mu	&=	\varphi.
		\label{prescription II}
}
%In the case of SM, 
Therefore, the effective potential is
\al{
V
	&=	{\lambda_\text{eff}(\mu)\over4}\varphi^4,
}
with the scale \eqref{prescription I} for the prescription I and scale \eqref{prescription II} for the prescription II,
where $\lambda_\text{eff}(\mu)$ in the SM is given by Eq.~\eqref{effective coupling}.

%%%%%%%%%%%%%%%%%%%%%%%%%%%%%%%%%%%%%%%%%%

\section{Cosmological constraints}
The overall normalization of the CMB fluctuation fixes~\cite{Ade:2013uln}
\al{\label{normalization}
A_s	&=	{V\over 24\pi^2\epsilon_VM_P^4}
	=	\paren{2.196^{+0.053}_{-0.058}}\times10^{-9},
%	=	2.2\times10^{-9}.
}
within $1\sigma$ CL.
Current Planck+WMAP bounds on the spectral index, its running, its running of running, and the tensor-to-scalar ratio are~\cite{Ade:2013uln}
\al{
n_s	&=	0.9514^{+0.0087}_{-0.0090},	&
{\df n_s\over\df\ln k}
	&=	0.001^{+0.016}_{-0.014},	&
{\df^2 n_s\over{\df\ln k}^2}
	&=	0.020^{+0.016}_{-0.015},	&
r	&=	0	\quad\text{(assumed)},\nn
n_s	&=	0.9583\pm0.0081,	&
{\df n_s\over\df\ln k}
	&=	-0.021\pm0.012,	&
{\df^2 n_s\over{\df\ln k}^2}
	&=	0\quad\text{(assumed)},	&
r	&<	0.25\quad\text{($2\sigma$ CL)},
	\label{Planck+WP}
}
at the pivot scale $k_*=0.05\,\text{Mpc}^{-1}$, within 1$\sigma$ CL unless otherwise stated.
The BICEP2 experiment has reported an observation of $r$~\cite{Ade:2014xna}:
\al{
r= 0.20^{+0.07}_{-0.05},
	\label{BICEP2}
}
within $1\sigma$ CL.

It has been pointed out that the BICEP2 result may become consistent with $r=0$ because the foreground effect can be sizable~\cite{Mortonson:2014bja,Flauger:2014qra}.
We also note that by including isocurvature perturbation, the 95\% CL bound on $n_s$ is roughly loosened to~\cite{Kawasaki:2014fwa}
\al{
0.93\lesssim n_s \lesssim 0.99
}
and that by including sterile neutrinos, the allowed range is shifted to~\cite{Zhang:2014dxk}
\al{
0.95\lesssim n_s \lesssim 1.02.
}
Given above, we will plot our results within wider ranges than those in Eqs.~\eqref{Planck+WP} and \eqref{BICEP2}:
\al{
0.93
	&\leq n_s
	\leq 1.02,	\\
-0.05
	&\leq {\df n_s\over\df\ln k}
	\leq 0.05,	\\
0	&\leq r
	\leq	0.3.
}

\section{Higgs inflation from Standard Model criticality}
In this section, we start from the minimal set of coefficients~\eqref{minimal coefficients}, and later include the term $\lambda_6\varphi^6$.

We \purple{expand} the effective potential of the Higgs field $V_\text{eff}$ on the flat space-time background around its minimum as in Eq.~\eqref{expansion}:
\al{\label{running minimal potential}
V
	&=	{\lambda_\text{eff}(\mu)\over4}\varphi^4,	\\
\lambda_\text{eff}(\mu)
	&=	\lambda_\text{min}+\sum_{n=2}^\infty{\beta_n\over\paren{16\pi^2}^n}\paren{\ln{\mu\over\mu_\text{min}}}^2.
\label{running effective coupling}
}
The choice of scale \eqref{prescription I} and \eqref{prescription II} \purple{corresponds} to the prescription I and II, respectively. 
As in Section~\ref{SM}, we can safely neglect the higher order terms with $n\geq3$, and we continue to omit them.
%In the potential analysis around the inflation scale, we can safely neglect the higher order terms with $n\geq3$, and we will omit them hereafter in this section.
 %The higher order terms in Eq.~\eqref{higher order potential} are ignored for the moment.

\subsection{Prescription I}\label{Prescription I}
\subsubsection{\magenta{Analysis in prescription I}}

\begin{figure}[tn]
\begin{center}
\hfill
\includegraphics[width=.6\textwidth]{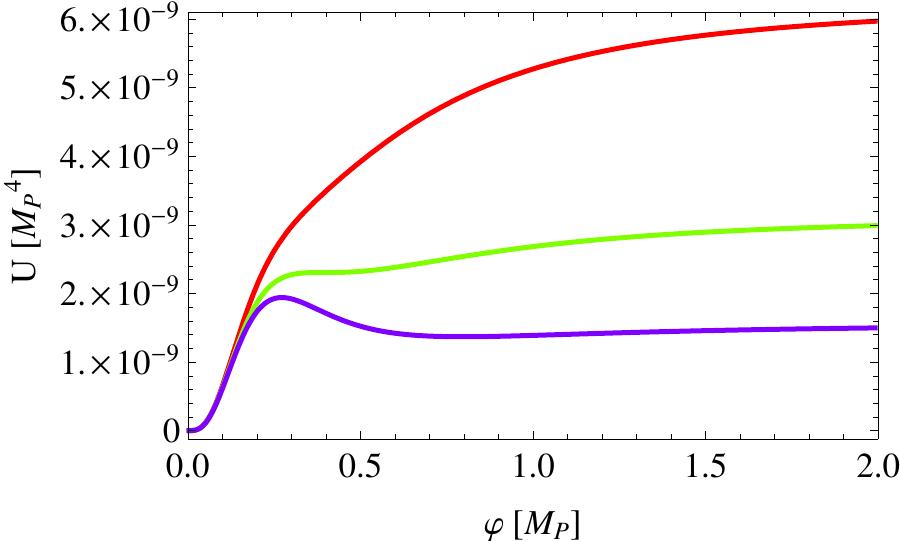}
\hfill\mbox{}\\
\caption{SM Higgs potential in the prescription I with $\xi=10$ and $c=1$, corresponding to $\mu_\text{min}=7.6\times10^{17}\GeV$, and with $\beta_2=0.5$. 
The red (upper), green (center) and purple (lower) lines are drawn with $\lambda_\text{min}=2\lambda_c$, $\lambda_c$, and $\lambda_c/2$, respectively.
The values of $\lambda_\text{min}=2\lambda_c$ and $\lambda_c/2$ are chosen just for illustration. 
Each line roughly corresponds to the one with the same color in Fig.~\ref{SMpotential}.
}
\label{potential}
\end{center}
\end{figure}

In the prescription I, the Higgs potential is given by \purple{Eqs.~\eqref{U potential} and \eqref{running minimal potential}} with the scale~\eqref{prescription I}.
%The potential is monotonically increasing function if $U'(\varphi)=\df U/\df\varphi$ is always positive. 
Concretely,
\al{
U(\varphi)
	&=
\frac{\varphi^4}{4(1+\xi \varphi^2/M_P^2)^2}\br{
	\lambda_\text{min}
	+\frac{\beta_2}{\paren{16\pi^2}^2}\sqbr{\ln\fn{\frac{1}{c}\sqrt{\frac{\xi \varphi^2/M_P^2}{1+\xi \varphi^2/M_P^2}}}}^2
	}
\nn
U'(\varphi)
	&=	\frac{\varphi^3M_P^6}{(M_P^2+\xi \varphi^2)^3}\br{
			\lambda_\text{min}
			+{\beta_2\over2\paren{16\pi^2}^2}\sqbr{
				1+2\ln\fn{\frac{1}{c}\sqrt{\frac{\xi \varphi^2}{M_P^2+\xi \varphi^2}}}
				}
			\ln\fn{\frac{1}{c}\sqrt{\frac{\xi \varphi^2}{M_P^2+\xi \varphi^2}}}
			},
}
where we define $c$ by
\al{
%h&\simeq \frac{M_P}{\sqrt{\xi}}\exp\left(\frac{\chi}{\sqrt{6}M_P}\right),\nn
%\simeq \frac{M_P}{\sqrt{\xi}} \frac{1}{\sqrt{1+e^{-2\chi/\sqrt{6}M_P}}}
%\simeq \frac{M_P}{\sqrt{\xi}} \left(1-\frac{1}{2}e^{-2\chi/\sqrt{6}M_P}\right)
\mu_\text{min}&=c\frac{M_P}{\sqrt{\xi}}.
	\label{c defined}
}
Note that we have defined $\mu_\text{min}$ to give the minimum of the effective coupling $\lambda_\text{eff}(\mu_\text{min})=\lambda_\text{min}$ on the flat spacetime background Eq.~\eqref{running effective coupling}.
%, and continue to use that definition in the curved background.
The stationary points $U'(\varphi_1)=0$ are given by
\al{
\varphi_1
	&=	{cM_P\over\sqrt{\xi}}{1\over \paren{e^{{1\over2}\sqbr{1\pm\sqrt{1-{\lambda_\text{min}\over\lambda_c}}}}-c^2}^{1/2}}.
		\label{phi1}
}
We can see the following:
\begin{itemize}
\item When $\lambda_\text{min}>\lambda_c$, the potential is a monotonically increasing function of $\varphi$.
	This case corresponds to the red (upper) line in Fig.~\ref{potential}.
\item When $\lambda_\text{min}=\lambda_c$:
	\begin{itemize}
	\item For $c\geq e^{1/4}$, the potential is monotonically increasing.
	\item For $c<e^{1/4}$, the potential has a stationary point at
		\al{
		\varphi_c
			=	{cM_P\over\sqrt{\xi}}{1\over\paren{\sqrt{e}-c^2}^{1/2}}.
		}
		In this case, $\varphi_c$ becomes a saddle point: $U'(\varphi_c)=U''(\varphi_c)=0$.
		This case corresponds to the green (center) line in Fig.~\ref{potential}.
	\end{itemize}
\item When $\lambda_\text{min}<\lambda_c$, we define $c_\pm:=\exp{{1\pm\sqrt{1-{\lambda_\text{min}\over\lambda_c}}}\over4}$:
	\begin{itemize}
	\item For $c\geq c_+$, the potential is monotonically increasing.
	\item For $c_-< c<c_+$, the potential has a stationary point given by the plus sign of Eq.~\eqref{phi1}.
	\item For $c\leq c_-$, the potential has two stationary points given by Eq.~\eqref{phi1}.
		This case corresponds to the purple (lower) line in Fig.~\ref{potential}.
	\end{itemize}
\end{itemize}

In this paper, we pursue the possibility that $\lambda(\mu_\text{min})\simeq0$ is realized by a principle beyond the ordinary local field theory, such as the multiple point criticality principle~\cite{Froggatt:1995rt,Froggatt:2001pa,Nielsen:2012pu}, classical conformality~\cite{Meissner:2006zh,Foot:2007iy,Meissner:2007xv,Iso:2009ss,Iso:2009nw,Aoki:2012xs,Iso:2012jn,Hashimoto:2013hta,Hashimoto:2014ela}, asymptotic safety~\cite{Shaposhnikov:2009pv}, the hidden duality and symmetry~\cite{Kawamura:2013kua,Kawamura:2013xwa}, and the maximum entropy principle~\cite{Kawai:2011qb,Kawai:2013wwa,Hamada:2014ofa,Kawana:2014vra}.

In practice, this amounts to the following:
$\mu_\text{min}$ is fixed for a given set of $M_H$ and $\alpha_s(M_Z)$ in the SM.
For a given $\mu_\text{min}$, we require $\xi$ to sit in
\al{
\xi	=	c^2{M_P^2\over\mu_\text{min}^2}
	&<	\sqrt{e}{M_P^2\over\mu_\text{min}^2}.
	\label{xi range}
}
That is, we consider the case $c<e^{1/4}$.
By tuning the top quark mass, we can always choose a $\lambda_\text{min}$ that is very close but larger than $\lambda_c$
so that we realize $U'(\varphi)\ll U(\varphi)/M_P$ and $U''(\varphi)\ll U(\varphi)/M_P^2$ around $\varphi\simeq\varphi_c$.\footnote{
More precisely, we need $U_{\chi}\ll U/M_P$ and $U_{\chi\chi}\ll U/M_P^2$, which are satisfied when $U_\varphi\ll U/M_P$ and $U_{\varphi\varphi}\ll U/M_P^2$ because we have $\df\varphi/\df\chi\sim1/\sqrt{6\xi}$ during the inflation.
}
%In Fig.~\ref{potential}, our choice corresponds to a red (upper) line but is very close to the green (middle) one.
\blue{
In Fig.~\ref{potential}, our choice is very close but slightly above the green (middle) line. %but corresponds to a red (upper) line.
}

In extensions of the SM, $\mu_\text{min}$ depends on newly-added parameters too.
Anyway we require Eq.~\eqref{xi range}, and choose a $\lambda_\text{min}$ that is very close to $\lambda_c$, with $\lambda_\text{min}>\lambda_c$, by the tuning of the top mass and possible other parameters.

We also need to consider the effect of the running of $\xi$~\cite{Buchbinder:1992rb,Odintsov:1990mt,Muta:1991mw,Yoon:1996yr}. However, this effect is small. 
More concretely, if $\xi$ and $\varphi$ are sufficiently large, 
$\xi$ is given around $\mu=\mu_\text{min}$ by~\cite{Bezrukov:2009db}
\al{
\xi(\mu)
	&\simeq	\xi_0\left\{1-\paren{\frac{3}{2}g_Y^2+3g_2^2-6y_t^2}\frac{1}{16\pi^2}\ln\frac{\mu}{\mu_\text{min}}\right\}
\simeq
\xi_0\left\{1+0.001\ln\frac{\mu}{\mu_\text{min}}\right\}
, &
%\beta_{\xi0}
	%&=	\xi\right|_{\mu=\mu_\text{min}}.
}
%Since %we consider the case where $\xi$ is larger than one, 
%the running of $\xi$ is not so important, and 
We treat $\xi$ as a constant in this paper.

\subsubsection{\magenta{Results in prescription I}}
The Higgs potential is determined by three parameters, $\xi$, $c$ and $\lambda_\text{min}$.
Qualitatively, $\xi$ determines the total suppression of the potential above the scale $\phi\gtrsim M_P/\sqrt{\xi}$, $c$ determines the maximum value of $\epsilon_V$ above the almost-saddle point, and $\lambda_\text{min}$ determines the number of $e$-folding.
%\al{
%U&=
%\epsilon&=
%}
We choose $\lambda_\text{min}$ such that we can have sufficient $e$-folding $N=60$.
For a fixed \blue{$A_s=2.2\times 10^{-9}$}, other cosmological parameters $n_s, r$ and $\df n_s/\df\ln k$ can be calculated as functions of $\xi$ and $c$.

\begin{figure}[tn]
\begin{center}
\hfill
\includegraphics[width=.45\textwidth]{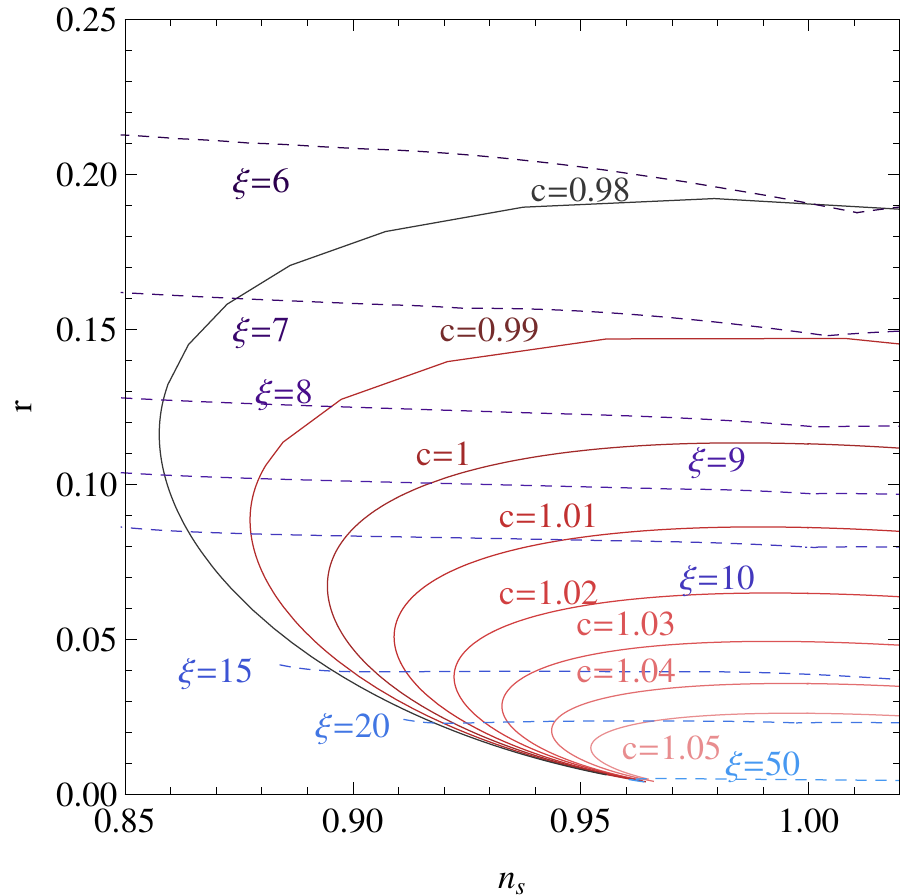}
\hfill
\includegraphics[width=.49\textwidth]{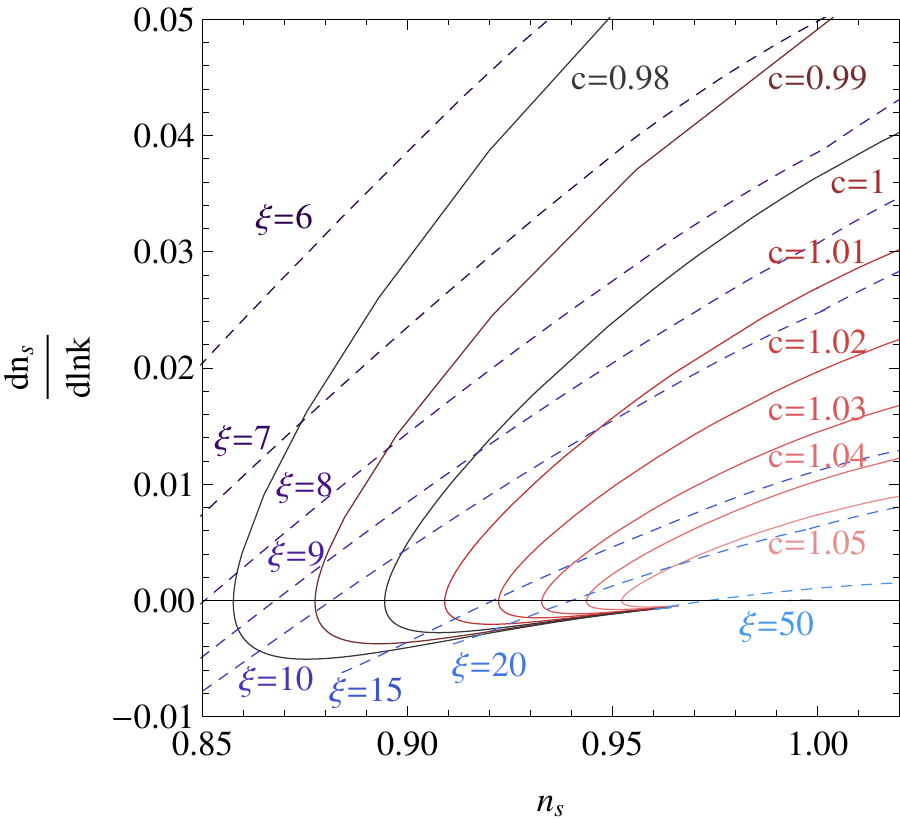}
\hfill\mbox{}\\
\caption{Left: $r$ vs $n_s$. Right: $\df n_s/\df \ln k$ vs $n_s$. 
\magenta{The solid and dashed contours are for fixed $c$ and $\xi$, respectively.}
\magenta{The left end of each dashed line for $\xi=15$, 20 and 50 corresponds to $c=0.94$.}
\magenta{The lower end of each solid line corresponds to $\xi=50$.}
%$c$ and $\xi$ vary within the region $0.94\leq c$ and $\xi\leq 50$.
}
\label{prediction_without_C6}
\end{center}
\end{figure}

We show the typical predictions of this model in Fig.~\ref{prediction_without_C6}.
\magenta{Each solid line corresponds to a constant~$c$. 
Dashed lines \purple{correspond} to the values of $\xi$ from 6 to 50 as indicated in the figure.}
In Fig.~\ref{prediction_without_C6}, we see that there is a minimum value of $\xi$ that can result in $r\lesssim 0.2$, namely $\xi_\text{min}\sim7$.
The model can reproduce $r=O(10^{-3})\sim 0.2$ and $n_s=0.9\sim1.0$.
These predictions are consistent with Planck or BICEP2 result~\cite{Ade:2013uln,Ade:2014xna}.
However, the value of $\df n_s/\df\ln k$ is slightly large.
The prediction is  $\df n_s/\df\ln k=O(0.01)$ for $r\gtrsim 0.05$.
\bblue{In Section~\ref{sixth term}, we will see that the \purple{introduction of small coefficients of} higher dimensional operator\purple{s may} ameliorate the situation.}

\begin{figure}[tn]
\begin{center}
\hfill
\includegraphics[width=.49\textwidth]{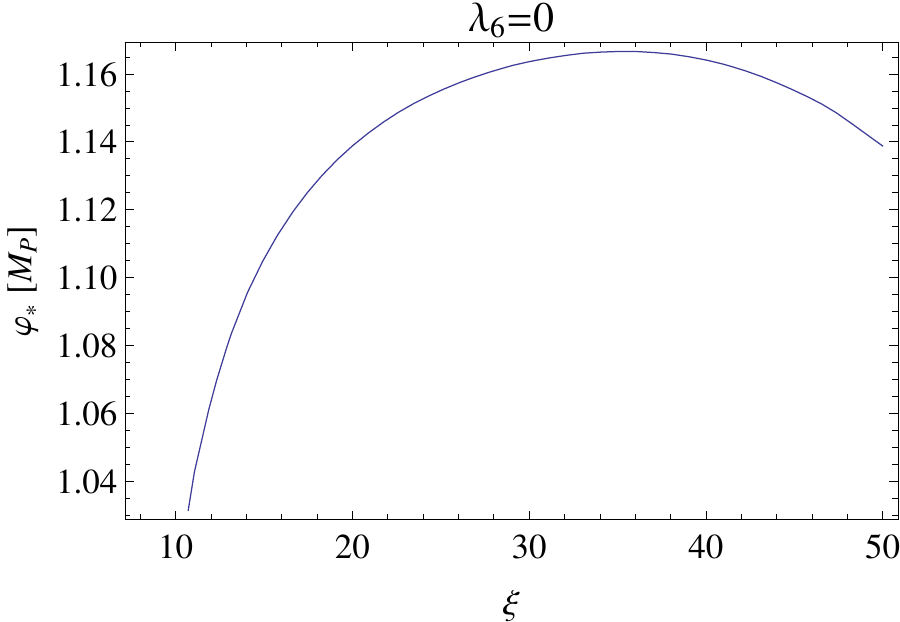}
\hfill
\includegraphics[width=.49\textwidth]{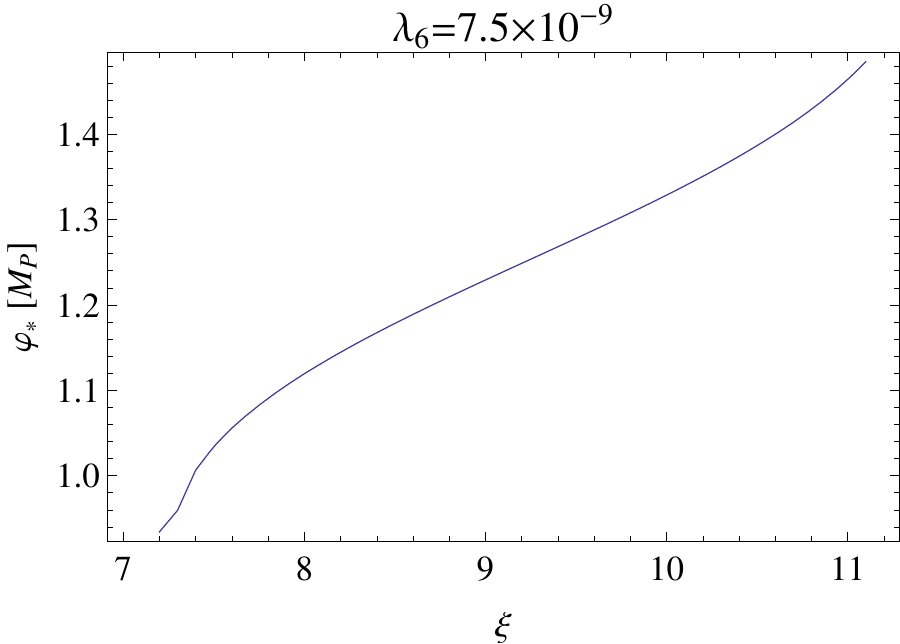}
\hfill\mbox{}\\
\caption{\magenta{Left and right: $\varphi_*$ as a function of $\xi$, with $\lambda_6=0$ and $7.5\times 10^{-9}$, respectively. Other parameters are taken as $c=1$ and $\beta_2=0.5$.}
}
\label{varphistar}
\end{center}
\end{figure}

Finally, we discuss the field value $\varphi_*$ that corresponds to the observed CMB fluctuation.
The left panel of Fig.~\ref{varphistar} shows $\varphi_*$ in the case of $c=1$ and $\beta_2=0.5$.\footnote{
Precisely speaking, there are two $\varphi_*$ which satisfies Eq.~\eqref{normalization} given $c, \xi$.
We plot the one solution which gives more desirable predictions on cosmological parameters, \magenta{namely, $n_s\lesssim 0.99$.}
}
We see that $\varphi_*$ is around the Planck scale: $\varphi_*\sim M_P$.

\subsubsection{\pink{Constraint on $\mu_\text{min}$}}\label{constraint on mu_min}
%Another interpretation is Multiple Point Principle.
%MPP requires $C_6$ is small in order not to disturb the flat potential around $10^{17\text{--}18}$GeV.
The above analysis shows the existence of the lowest possible value of $\xi$, which is $\xi_\text{min}\sim 7$.
It is a necessary condition that \magenta{$\mu_\text{min}$, which is obtained from the parameters at low energy, satisfies} $\mu_\text{min}\lesssim M_P/\sqrt{\xi_\text{min}}$ for \magenta{any} successful Higgs inflation \magenta{with $\xi>\xi_\text{min}$}. 
However, as we have observed in Sec.~\ref{SM}, SM one-loop effective potential takes its minimum above $M_P/\sqrt{\xi_\text{min}}$ although the tree level potential can realize $\mu_\text{min}\lesssim M_P/\sqrt{\xi_\text{min}}$.
It appears that it is difficult to do our Higgs inflation in SM. 

However, taking into account the ambiguity coming from non-renormalizable non-minimal coupling $\xi$, there still remains a possibility of realizing $\mu_\text{min}\lesssim M_P/\sqrt{\xi_\text{min}}$\purple{~\cite{Bezrukov:2014bra}}.
\magenta{Around the scale $M_P/\xi$, we match $\lambda$ in the SM without $\xi$ and $\lambda_\xi$ in the SM with $\xi$:
\al{
\lambda=\lambda_\xi+(\text{threshold corrections}),
}
where the threshold corrections generally contain power divergences and cannot be determined unless we specify a UV theory beyond the cutoff.
One expects that the threshold corrections start from one loop order.
Because they are of the same order as the difference between the tree and one-loop effective potentials, it may result in $\mu_\text{min}\lesssim M_P/\sqrt{\xi_\text{min}}$.
See also the similar discussion in %Section~\ref{SM} 
footnote \ref{gauge dependence} regarding the gauge dependence.

In Section~\ref{DM section}, we will see that we can easily obtain $\mu_\text{min}\lesssim M_P/\sqrt{\xi_\text{min}}$ in the Higgs portal scalar DM model without referring to such arguments.}
\pink{Argument of this section applies also to the prescription II shown in Sec.~\ref{prescription II section}.}

\subsubsection{\purple{Estimation of the effects of higher dimensional operators in prescription I}}\label{sixth term}

\begin{figure}[tn]
\begin{center}
\hfill
\includegraphics[width=.45\textwidth]{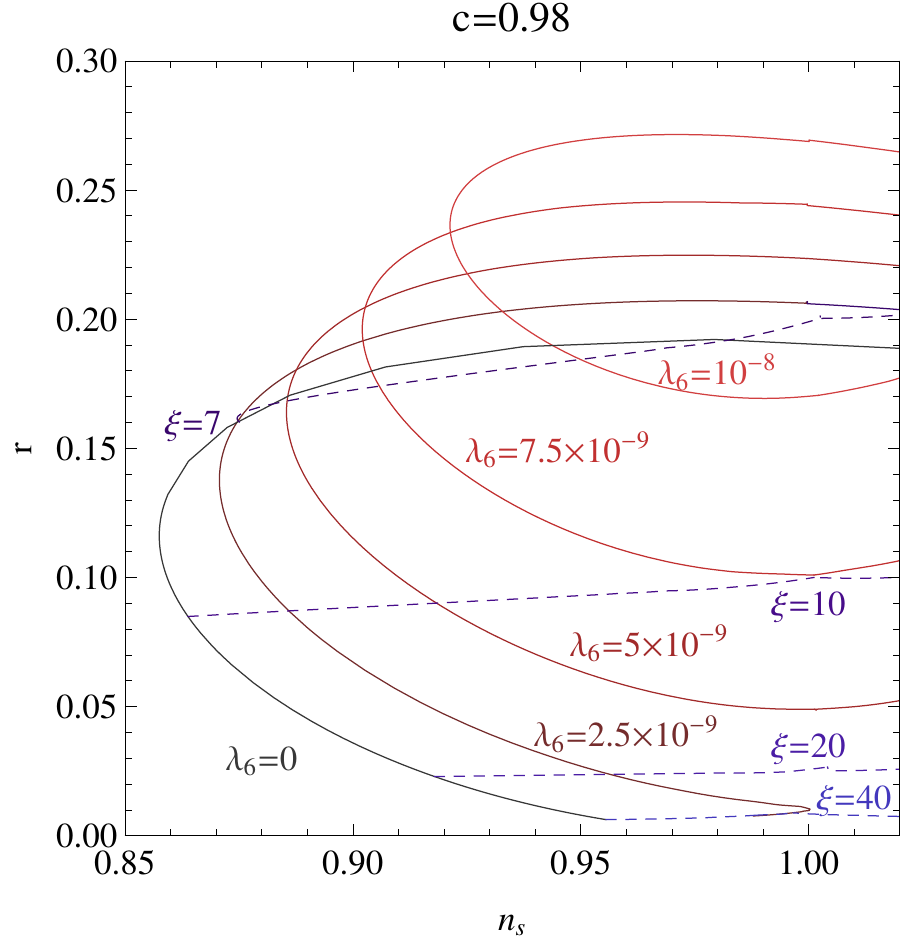}
\hfill
\includegraphics[width=.45\textwidth]{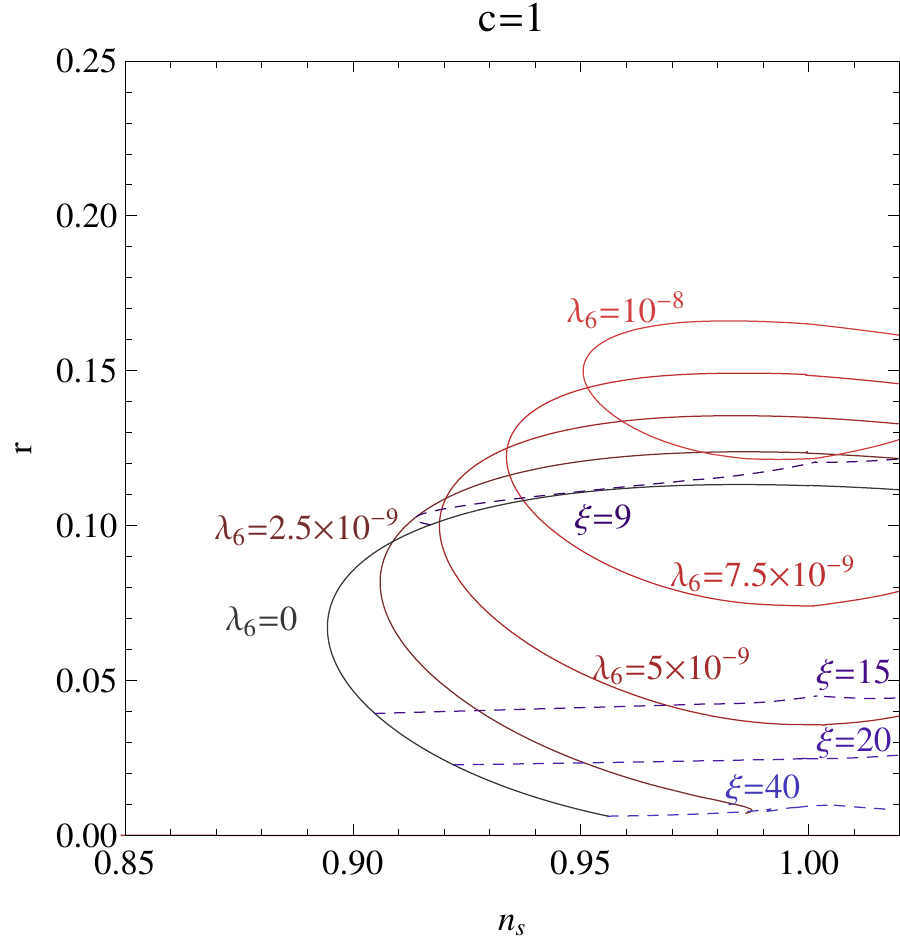}
\hfill\mbox{}\\
\caption{\magenta{$r$ vs $n_s$, with $c=0.98$ (left) and 1 (right). Solid and dashed contours are for fixed $\lambda_6$ and $\xi$, respectively. We put $\beta_2=0.5$.}}
\label{r vs ns}
\end{center}
\end{figure}

\begin{figure}[tn]
\begin{center}
\hfill
\includegraphics[width=.49\textwidth]{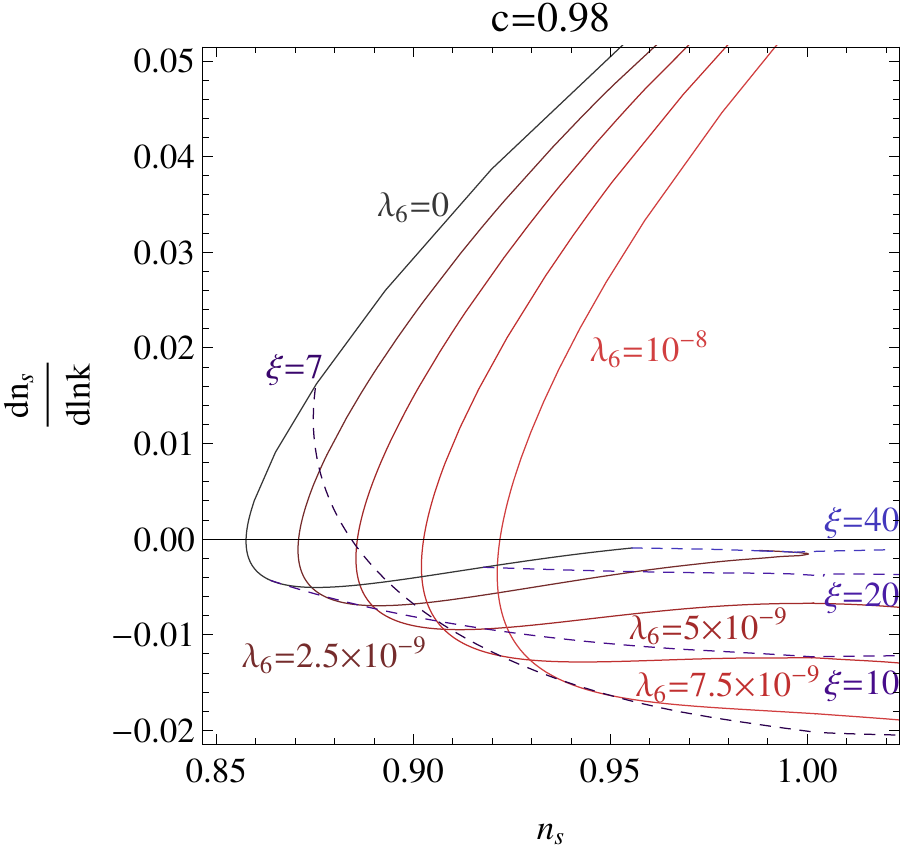}
\hfill
\includegraphics[width=.49\textwidth]{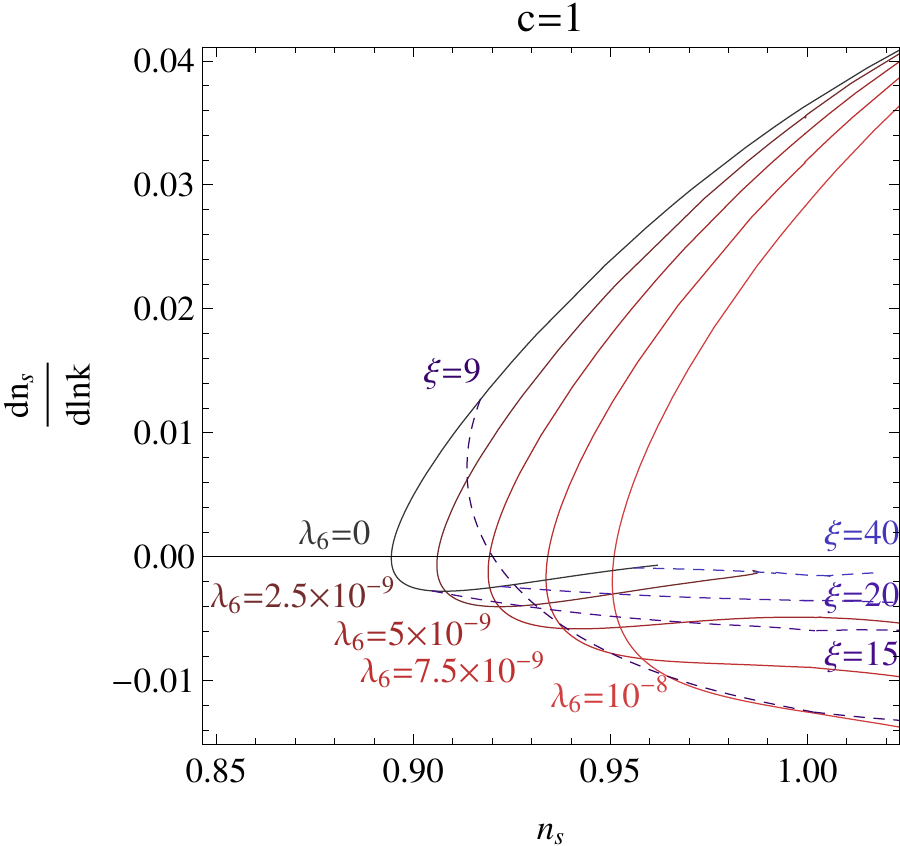}
\hfill\mbox{}\\
\caption{\magenta{$\df n_s/\df \ln k$ vs $n_s$, with $c=0.98$ (left) and 1 (right). Solid and dashed contours are for fixed $\lambda_6$ and $\xi$, respectively. We put $\beta_2=0.5$.}}
\label{dnsdlnk vs r}
\end{center}
\end{figure}

%\rred{As we take the SM as an effective theory below the Planck scale, higher dimensional operators beyond the dimension-four terms appear at higher scales, and their actual sizes are not predictable without knowing the concrete UV completion.}
%\pink{In this scenario, we have assumed that several lowest levels of the Planck-suppressed higher dimensional operators are much smaller than a ``natural'' value of order unity, which is typically expected in the effective field theory of a strongly-coupled ultraviolet dyanamics.}
\pink{
As we have seen in the previous sections, the extrapolation of the low energy data shows that the Higgs potential in the SM is flat around the string scale. 
\purple{This flatness can be broken if we introduce arbitrary strengths to the higher dimensional Planck-suppressed operators.}
%We take this fact as an indication that the UV completion indeed makes the Higgs potential flat around the Planck scale, where we can no longer handle it by a local quantum field theory.
In order to \purple{examine the effects of such operators on} the cosmological data, we \purple{consider, for example, a} small sixth order term in the Jordan frame
\al{
\Delta V=\lambda_6 \frac{\varphi^6}{M_P^2}.
	\label{dimension six}
}}
%as \purple{an example.}
%and examine its impact on $\df n_s/\df\ln k$.
\purple{Here we do not claim that Eq.~\eqref{dimension six} gives the leading contribution among the higher dimensional operators,
but simply estimate the ambiguity in the value of $\df n_s/\df \ln k$ discussed in the previous section.
Of course, we can give any form to $\df n_s/\df \ln k$ as a function of $k$ if we introduce arbitrary strengths to the higher dimensional operators. As we will see below, the single term~\eqref{dimension six} allows the value of $\df n_s/\df \ln k$ at $k_*$ to reside in the favored region. However, this should not be taken as a prediction of the value of $\lambda_6$ but as an estimation for the allowed magnitude of the coefficients of the higher dimensional operators.
}

In the Einstein frame potential, the term~\eqref{dimension six} becomes,
\al{
\Delta U=\lambda_6 \frac{\varphi^6}{\left(1+\xi \varphi^2/M_P^2\right)^2}.
	\label{sixth order term}
}
%In this paper, we have assumed that the Higgs potential around the Planck scale becomes flat based on the extrapolation of the low energy data. We take this fact as an indication that the UV completion indeed makes the Higgs potential flat around the Planck scale, where we can no longer handle the Higgs potential by a local quantum field theory. 
%In terms of the low energy effective field theory, this is e.g.\ realized by the parameter set~\eqref{minimal coefficients}. In general we expect that there are small deviations of these couplings from zero.
%
%\pink{Obviously, we can always fit $\df n_s/\df \ln k$ by adding arbitrary number of Planck suppressed operators.}
%\rred{Here we examine the lowest higher dimensional operator in the Jordan frame as a simplest example to know the degree of ambiguity, within which we can easily fit $\df n_s/\df \ln k$ to the observed value.}
%
\magenta{In Figs.~\ref{r vs ns} and \ref{dnsdlnk vs r}, we plot the contours for fixed $\lambda_6\leq 10^{-8}$ with the solid lines, in the $r$ vs $n_s$ plane and the $\df n_s/\df\ln k$ vs $n_s$ one, respectively.
We also plot the contours for fixed $\xi$ and $\lambda_6$ in the dashed and solid lines, respectively.}
We can realize the $r\simeq0.1$, $n_s\simeq0.96$, and $\df n_s/\df\ln k\simeq-0.01$ simultaneously.
%It is quite interesting that the higher moment $\df n_s/\df\ln k$ (and further $\df^2n_s/\df\ln k^2$, \dots; $\df n_t/\df\ln k$, \dots \magenta{being calculable exactly the same way}), which will be explored in future experiments, can constrain such a Planck-suppressed operator.
Finally, $\varphi_*$ has been plotted in the right panel of Fig.~\ref{varphistar} with $\lambda_6=5\times 10^{-9}$, $c=1$, and $\beta_2=0.5$.

\purple{The other higher dimensional operators should also have the coefficients $\lesssim \mathcal O(10^{-8})$ in order to keep the flatness of the potential.
Their} smallness \purple{may} be understood for example as a tiny explicit breaking of \pink{the} asymptotic scale invariance in Jordan frame (\pink{the} shift symmetry in Einstein frame)~\cite{Bezrukov:2013fka}.

%%%%%%%%%%%%%%%%%%%%%%%%%%%%%%%%%%%%%%%%

\subsection{Prescription II}\label{prescription II section}
\subsubsection{\magenta{Analysis in prescription II}}
In the prescription II, the Higgs potential is given by Eqs.~\eqref{U potential} and \eqref{running minimal potential} with $\mu=\varphi$,
\al{
U(\varphi)&=\frac{\lambda(\varphi)}{4}\frac{\varphi^4}{\paren{1+\xi \varphi^2/M_P^2}^2},
}
which gives
\al{
U
	&=	\frac{X^4}{(1+X^2)^2}\frac{1}{4}\left(\lambda_\text{min}+\frac{\beta_2}{\paren{16\pi^2}^2}\left(\ln\frac{X}{c}\right)^2\right)\paren{M_P\over\sqrt{\xi}}^4,\\
U_\varphi
	&=	{X^3\over\paren{1+X^2}^3}
			\br{
				\lambda_\text{min}
				+{\beta_2\over2\paren{16\pi^2}^2}\paren{1+X^2}\ln{X\over c}
				+{\beta_2\over\paren{16\pi^2}^2}\paren{\ln{X\over c}}^2
				}\paren{M_P\over\sqrt{\xi}}^3,	\label{Uprime}
}
where $X=\displaystyle{\varphi\over M_P/\sqrt{\xi}}$.
Then the slow-roll parameter~\eqref{epsilon defined} becomes
\al{
\epsilon_V
	&=
{8\xi\over X^2+\paren{1+6\xi}X^4}
{\paren{
	\lambda_\text{min}+\frac{\beta_2}{2\paren{16\pi^2}^2}\ln{X\over c}\paren{1+X^2+2\ln{X\over c}}
	}^2
	\over
	\paren{\lambda_\text{min}+\frac{\beta_2}{\paren{16\pi^2}^2}\paren{\ln{X\over c}}^2}^2}.
}
For $\varphi\gg M_P/\sqrt{6}\xi$ and $\xi\gg1/6$, we obtain
\al{
\epsilon_V
	&\simeq
\frac{4}{3X^4}
\frac{\left(\lambda_\text{min}+\frac{\beta_2}{2\paren{16\pi^2}^2}\ln\frac{X}{c}\paren{1+X^2+2\ln\frac{X}{c}}\right)^2}{\left(\lambda_\text{min}+\frac{\beta_2}{\paren{16\pi^2}^2}\left(\ln\frac{X}{c}\right)^2\right)^2}.}
Similarly we have
\al{
\eta_V
	&\simeq {4\over3X^4}\paren{1-X^2+\frac{\beta_2}{4\paren{16\pi^2}^2}
				\frac{\paren{1+X^2}\paren{1+X^2+6\ln\frac{X}{c}}}{\lambda_\text{min}+\frac{\beta_2}{\paren{16\pi^2}^2}\left(\ln\frac{X}{c}\right)^2}}.
}
These expressions are in agreement with those in the original Higgs inflation~\cite{Bezrukov:2007ep} if we take $\beta_2=0$ and $X\gg1$.

The $e$-folding $N$ is written by
\al{
N=\int^{X_*}_{X_\text{end}} \frac{\df\chi}{M_P} \frac{1}{\sqrt{2\epsilon_V}}
=\int \frac{\df\chi}{\df X}\frac{\df X}{\sqrt{2\epsilon_V}}\frac{1}{M_P},
}
where
\al{
\frac{\df\chi}{\df X}
	&=\frac{\sqrt{1+\paren{1+6\xi}X^2}}{1+X^2}\frac{M_P}{\sqrt{\xi}}
	\simeq \frac{\sqrt{6X}}{1+X^2}M_P.
}
In the last step, we \purple{have} used the same limit as above: $X\gg1/\sqrt{6}\xi$ and $\xi\gg1/6$.
Finally, we can write $N$ as a function of $X$ in that limit:
\al{
N\simeq\int^{X_*}_{X_\text{end}} dX 
\frac{3X^3}{1+X^2}\frac{\lambda_\text{min}+\frac{\beta_2}{\paren{16\pi^2}^2}\left(\ln \frac{X}{c}\right)^2}
{2\lambda_\text{min}+\frac{\beta_2}{\paren{16\pi^2}^2}\ln\frac{X}{c}\left(1+X^2+2\ln\frac{X}{c}\right)}.
}
This is also in agreement with Ref.~\cite{Bezrukov:2007ep} if we put $\beta_2=0$ and $X_*\gg1$.

\subsubsection{\magenta{Results in prescription II}}

\begin{figure}[tn]
\begin{center}
\hfill
\includegraphics[width=.45\textwidth]{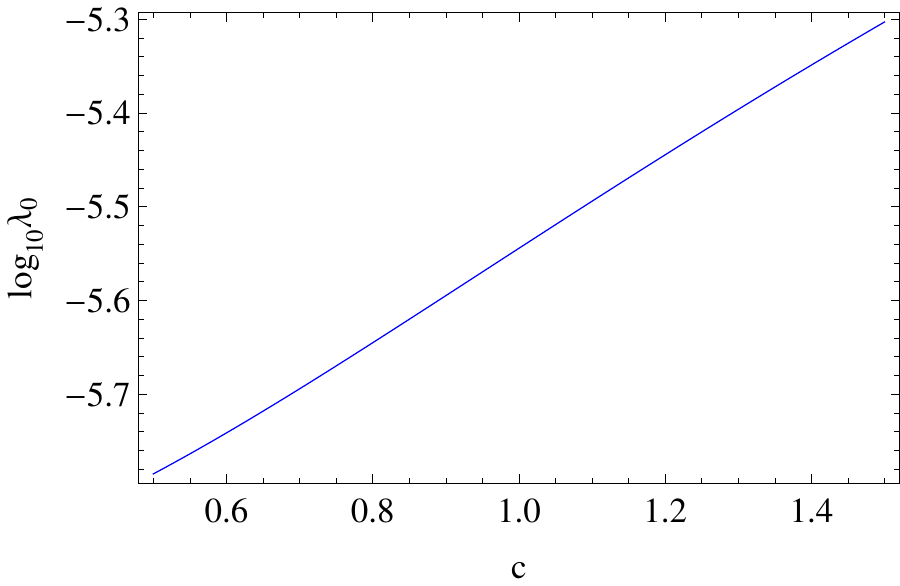}
\hfill
\includegraphics[width=.49\textwidth]{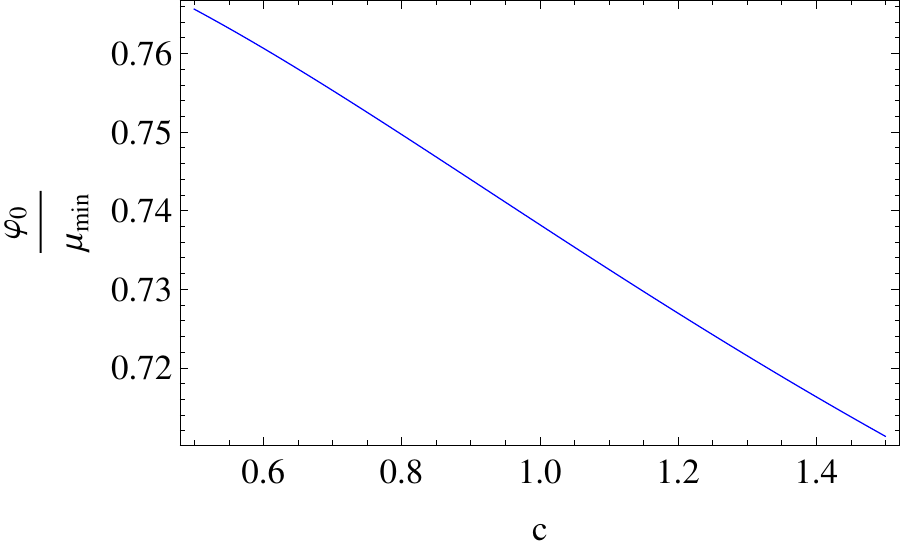}
\hfill\mbox{}\\
\caption{\blue{
Left: \magenta{$\lambda_0$,} the minimal value of $\lambda_\text{min}$ to maintain monotonicity of the potential, as a function of $c$.
Right: \magenta{$\varphi_0$,} the position of the saddle point \magenta{when we set $\lambda_\text{min}=\lambda_0$}, as a function of $c$. } 
}
\label{lambda_cplot}
\end{center}
\end{figure}

\magenta{Let us numerically estimate the lowest possible value of $\lambda_\text{min}$ that allows $U(\varphi)$ to be monotonically increasing.
We call this value $\lambda_0$.}
%Let us numerically estimate the condition on $\lambda_\text{min}$ such that $U(\varphi)$ becomes monotonically increasing.
%
%Let $\lambda_{0}$ denote the lowest possible value of $\lambda_\text{min}$ that makes $U(\varphi)$ monotonically increasing.
%
\magenta{In the prescription I, such a value was $\lambda_c$, whereas in the prescription II, $\lambda_0$ is a function of} $\beta_2$ and $c$.
\magenta{Note that $\lambda_0$ is} independent of $\xi$ because the expression in the braces in Eq.~\eqref{Uprime} only depends on \magenta{$X$, and explicit dependence on $\xi$ drops out of it}.
In Fig.~\ref{lambda_cplot}, we plot $\lambda_{0}$ and the position of the saddle point $\varphi_0$ as functions of $c$ for a fixed $\beta_2=0.5$.
We see that $\lambda_0\sim10^{-5.5}$ and  $\varphi_0\simeq0.73\mu_\text{min}$ for $c=1$.

\begin{figure}[tn]
\begin{center}
\hfill
\includegraphics[width=.5\textwidth]{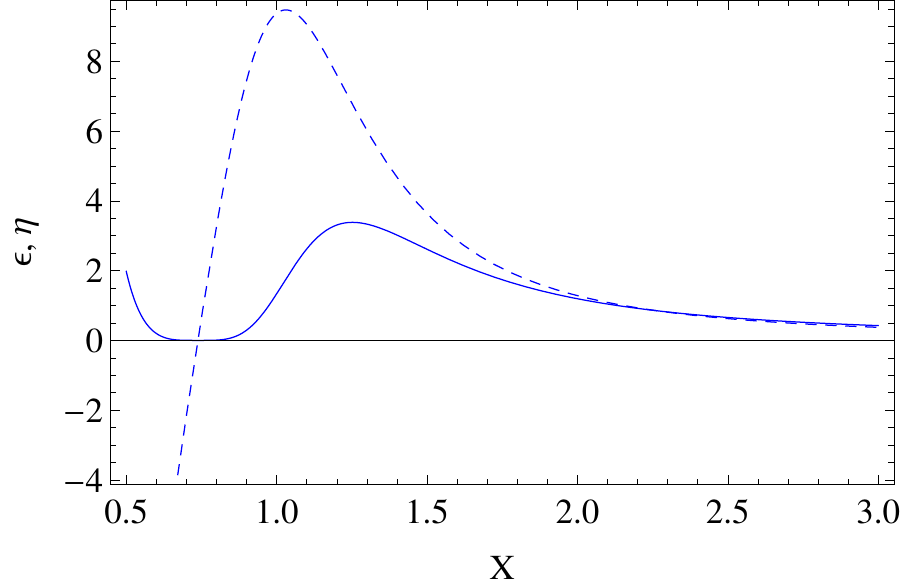}
%\hfill
%\includegraphics[width=.49\textwidth]{slowroll2.pdf}
\hfill\mbox{}\\
\caption{Slow roll parameters $\epsilon$ (solid) and $\eta$ (dashed) as functions of $X=\varphi/\paren{M_P/\sqrt{\xi}}$. We have set $c=1$, $\beta_2=0.5$, and $\lambda_\text{min}=\lambda_0$.}
\label{slowroll}
\end{center}
\end{figure}

The potential is determined by $\lambda_\text{min}$, $c$ and $\xi$.
To be specific, we consider the $c=1$ case hereafter.
We plot the $\epsilon_V$ in Fig~\ref{slowroll} with $c=1$, $\beta_2=0.5$, and $\lambda_\text{min}=\lambda_0$.
The solid and dashed lines represent $\epsilon_V$ and $\eta_V$, respectively.
We can see that $\epsilon_V\simeq \eta_V\simeq1$ around $X\simeq2$.
%Therefore we set $X$ corresponding to the end of inflation, $X_\text{end}$, to be  $X_\text{end}=2$.
\magenta{Therefore the end of inflation corresponds to $X_\text{end}\simeq 2$.}

\begin{figure}[tn]
\begin{center}
\hfill
\includegraphics[width=.5\textwidth]{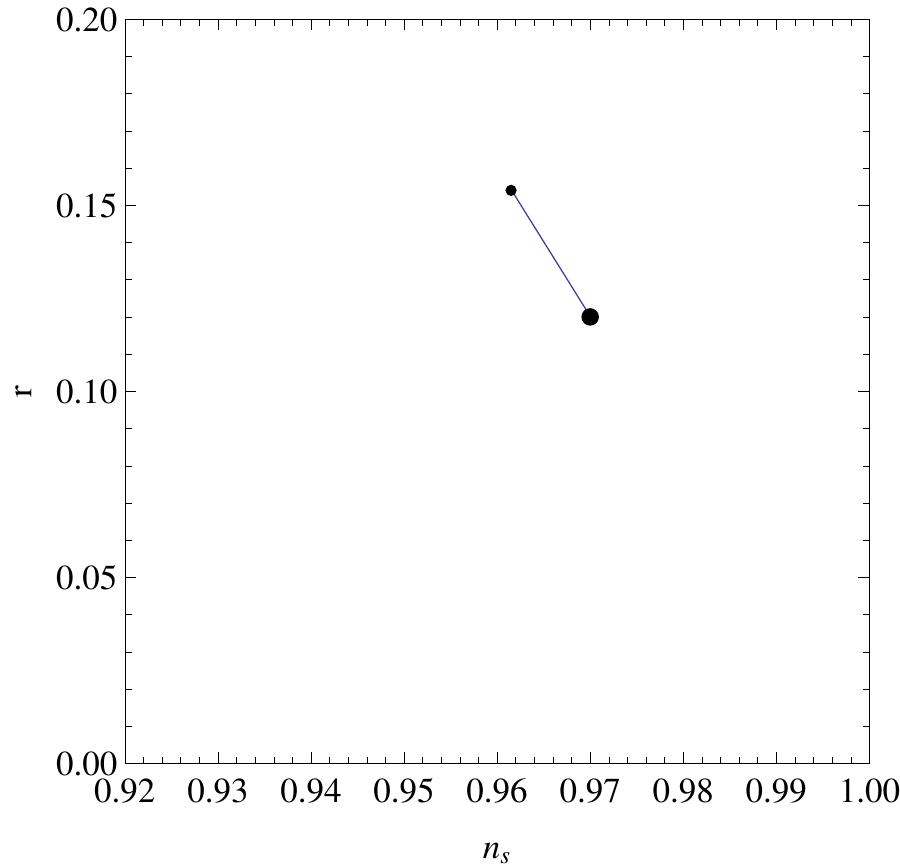}
\hfill\mbox{}\\
\caption{$n_s$ vs. $r$. The small and large dots represent $N_*=50$ and $65$.
}
\label{prescription II result}
\end{center}
\end{figure}

We can calculate the prediction of inflationary parameters with $c=1$, $\beta_2=0.5$, and $\lambda_\text{min}=\lambda_0$. 
$N=50$ and $65$ \purple{correspond} to $X_*\simeq360$ and $790$, respectively. 
We fix $\xi$ in such a way that Planck normalization is satisfied,
\al{
A_s=\frac{U}{24\pi^2 M_P^4 \epsilon_V}=2.2\times 10^{-9}.
}
By using this condition, $\xi$ becomes $190$ and $240$ for $N=50$ and 65, respectively.
The prediction of $n_s$ and $r$ is shown in Fig.~\ref{prescription II result}. 
$\df n_s/\df\ln k$ is small in this case, $\df n_s/\df\ln k\ll O(10^{-2})$.
These predictions are just \magenta{the same as the} chaotic inflation, as discussed in Ref.~\cite{Hamada:2014iga}. 

We note that the argument in this subsection implicitly assumes that Planck scale physics does not modify the Higgs potential above the UV cutoff. % which is computed using field theory.

\section{Scalar dark matter model}\label{DM section}
Next we consider the model which includes Higgs portal singlet scalar DM $S$ \red{\cite{SDM,Cline:2013gha}; see also Ref.~\cite{CSKim}.}
The Lagrangian is~\red{\cite{Zee}}
\al{\label{Lagrangian}
\mathcal{L}=\mathcal{L}_{\text{SM}}
+\frac{1}{2}(\partial_{\mu}S)^2-\frac{1}{2}m_S^2S^2
-\frac{\rho}{4!}S^4-\frac{\kappa}{2}S^2 H^\dagger H.
}
We put subscript $Z$ on the new parameters at the $Z$ mass scale $\mu=M_Z$, that is,
$\kappa_Z=\kappa\fn{\mu=M_Z}$, and $\rho_Z=\rho\fn{\mu=M_Z}$.
If we require perturbativity up to the cutoff scale, these parameters should be $\kappa_Z\lesssim0.4$ and $\rho_Z\lesssim 0.6$~\cite{Hamada:2014xka}.
The one-loop effective potential in this model is given by
\al{
V
	&=	V_\text{tree}+\Delta V_\text{1-loop,\,DM},
		\label{one-loop}
}
\al{
V_\text{tree}
&=
\blue{
e^{4\Gamma(\varphi)}
}
{\lambda(\mu)\over 4}\varphi^4,\nn
\Delta V_\text{1-loop,\,DM}
&=
\Delta V_\text{1-loop}
+
\frac{m_\text{DM}^4}{64\pi^2}\left(\ln\frac{m_\text{DM}(\varphi)^2}{\mu^2}-\frac{3}{2}\right),
}
where \blue{$m_\text{DM}(\varphi)=\sqrt{{\kappa\varphi^2\over2}e^{2\Gamma(\varphi)}+m_S^2}$. %and 
$\Delta V_\text{1-loop}$ and $\Gamma$ are given by SM one-loop potential~\eqref{1-loop lambda} and Eq.~\eqref{wavefunction}, respectively.}

\begin{figure}[tn]
\begin{center}
\hfill
\includegraphics[width=.32\textwidth]{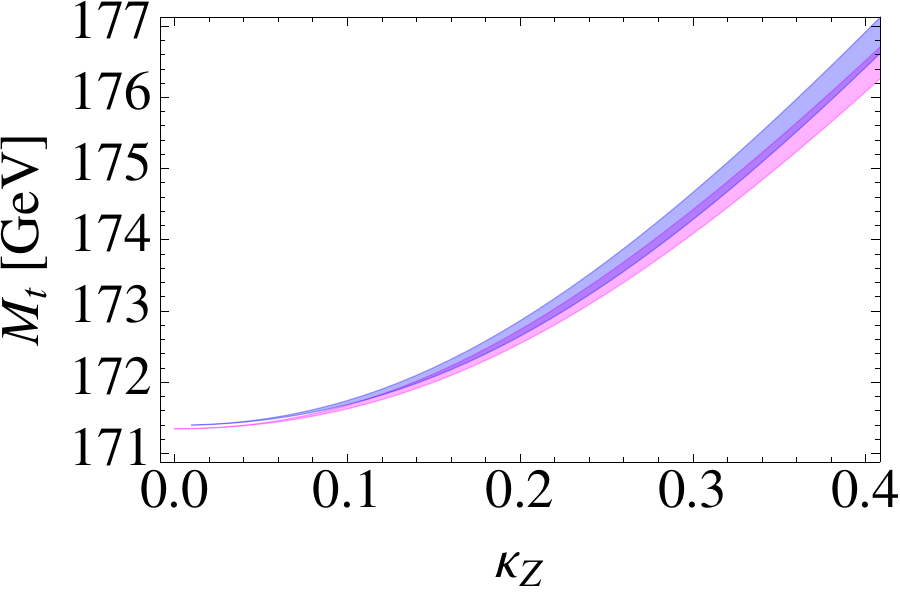}
\hfill
\includegraphics[width=.32\textwidth]{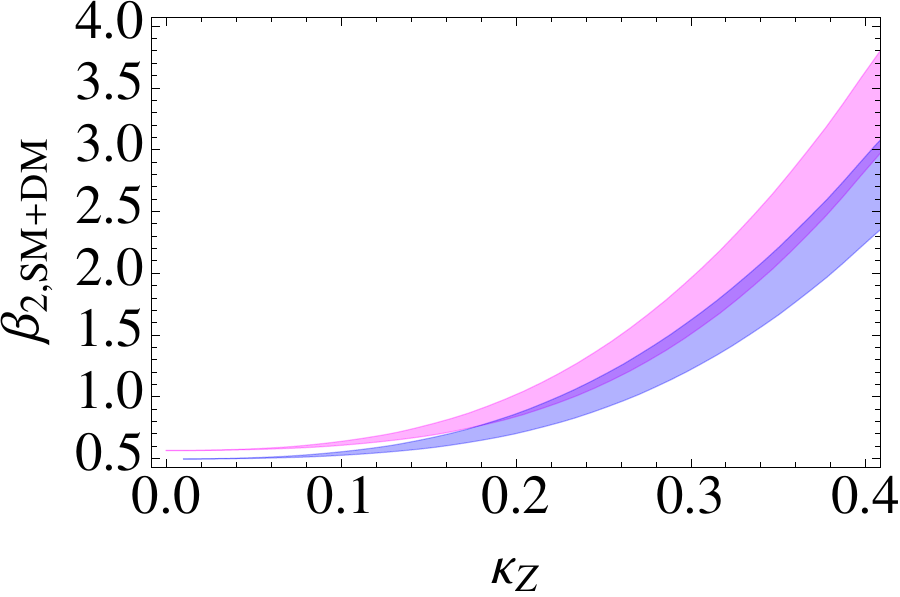}
%\hfill\mbox{}\\
\hfill
\includegraphics[width=.32\textwidth]{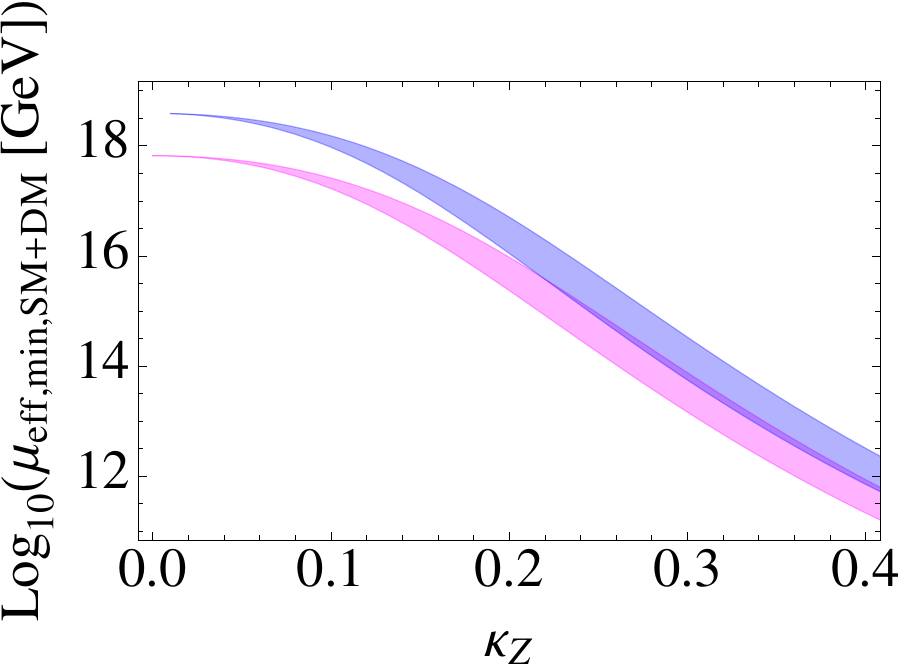}
%\\
%\hfill
%\includegraphics[width=.49\textwidth]{kappa0_vs_beta22.pdf}
%\hfill
%\includegraphics[width=.32\textwidth]{kappa0_vs_Mt2.pdf}
%\hfill
%\includegraphics[width=.32\textwidth]{kappa0_vs_beta22.pdf}
%\hfill\mbox{}\\
%\hfill
%\includegraphics[width=.32\textwidth]{kappa0_vs_mumin2.pdf}
\hfill\mbox{}\\
\caption{\magenta{$M_t$ (left), $\beta_2$ (center), and $\mu_\text{min}$ (right) are plotted as functions of $\kappa_Z$.
Red (lower) and blue (upper) bands correspond to the tree and 1-loop potentials, respectively.
The band width comes from the requirement of perturbativity of $\rho$ up to the string scale~\cite{Hamada:2014xka}: $0\leq\rho_Z\leq0.6$.}
\blue{
$M_H$ and $m_S$ are set to be $125.9$\,GeV and $0$, respectively.}
}
\label{scalar}
\end{center}
\end{figure}

We plot $M_t$, $\beta_2$, and $\mu_\text{min}$ as function\purple{s} of $\kappa_Z$ imposing the existence of the saddle point in Fig.~\ref{scalar}.
Here we use two loop RGEs~\cite{Hamada:2014xka} and put  $M_H=125.9\GeV, \alpha_s=0.1184$.
\magenta{The band width comes from the requirement of perturbativity of $\rho$ up to string scale~\cite{Hamada:2014xka}: $0\leq\rho_Z\leq0.6$.}
The red (lower) and blue (upper) bands correspond to the tree and one-loop effective potentials, respectively.
%赤はtree, 青はone-loopのeffective potentialを用いた場合である. 
%ここでRGEは2loopのものを用い~\cite{Hamada:2014xka}, Higgs massとstrong couplingはcentral valueに固定した, $M_H=125.9\GeV, \alpha_s=0.1184$.
\magenta{From this figure, we see that $\mu_\text{min}$ can become smaller than $M_P$ by adding $\kappa$ and that $\beta_2$ remains to be $O(1)$.}
\magenta{In particular, the addition of the scalar DM does not alter the existence of the minimum of $\lambda_\text{eff}(\mu)$, which is essential in this inflation scenario with criticality.}

%この図から適当な$\kappa$を加えることで$\mu_\text{min}$は$M_P$以下の任意の値をとることができ, $\beta_2$は$O(1)$の量をとることが読み取れる.
%この模型においてもHiggs inflation from SM criticalityは可能である.
%DMを加えた模型についてもsaddle pointが出来るときの種々のparameterの関係をFig.~\ref{scalareff}に示した. ただし, こちらでは$M_H=125.9\GeV, \alpha_s=0.1184$に固定している.

%\begin{figure}[tn]
%\begin{center}
%\hfill
%\includegraphics[width=.49\textwidth]{kappa0_vs_Mteff.pdf}
%\hfill
%\includegraphics[width=.49\textwidth]{kappa0_vs_beta2eff.pdf}
%\hfill\mbox{}\\
%\hfill
%\includegraphics[width=.49\textwidth]{kappa0_vs_mumineff.pdf}
%\hfill
%\includegraphics[width=.49\textwidth]{kappa0_vs_beta2eff2.pdf}
%\hfill\mbox{}\\
%\caption{The band corresponds to $\rho_Z$ uncertainty. $M_H$ is set to be $125.9$GeV.
%}
%\label{scalareff}
%\end{center}
%\end{figure}

\section{Summary}
We have considered the Higgs inflation model which contains non-minimal coupling $\xi \varphi^2 \mathcal{R}$~\cite{Bezrukov:2007ep}. 
\red{Conventional wisdom \purple{was} that a large non-minimal coupling $\xi\sim10^4$ is required to fit the COBE normalization, $\delta T/T\sim 10^{-5}$},  and cosmological predictions are $n_s=0.967$ and the small tensor-to-scalar ratio, $r=3\times 10^{-3}$.
In the letter~\cite{Hamada:2014iga}, we have reconsidered this model in light of the discovery of the Higgs boson, which indicates the criticality of the SM. \magenta{That is,} if \magenta{the SM parameters are} tuned so that the saddle point appears, it is possible to realize a Higgs inflation with moderate $\xi$ and generate $O(0.1)$ tensor to scalar ratio $r$. \magenta{The value of} $\xi$ is $O(10)$ for the prescription I and $O(100)$ for the prescription II. 

In this paper, we investigate the cosmological predictions of this Higgs inflation in \red{greater} detail.
\purple{To realize this Higgs inflation scenario,} it is \magenta{essential} that \red{the \magenta{effective} Higgs quartic coupling \magenta{$\lambda_\text{eff}$} takes} \magenta{its} minimum around the scale $10^{17\text{--}18}\GeV$.
\magenta{The Higgs potential around the inflation scale is determined by the position $\mu_\text{min}$ of the minimum of \magenta{$\lambda_\text{eff}$}, the minimum value $\lambda_\text{min}$, and the second derivative $\beta_2$ around the minimum, in addition to the non-minimal coupling $\xi$.}
We calculate the cosmological predictions as functions of above parameters.
We also \purple{estimate the effects of higher dimensional operators} by considering the $\lambda_6\varphi^6/M_P^2$ term \purple{as an example}.
\purple{We find that the coefficients of the higher dimensional operators should be as small as $10^{-8}$ in order to account for} the scalar and tensor perturbation\purple{s} consistent with the Planck \magenta{and} BICEP2 results.
Although we have \purple{concentrated} on the SM and the Higgs portal scalar DM model in this paper, \purple{one may apply our analysis} to \purple{the} other models beyond the SM \purple{by evaluating} $\lambda_\text{min}$, $\mu_\text{min}$, $\beta_2$ \purple{in terms of} the model parameters. 
%この論文ではSMと$Z_2$ scalar DMを含む拡張模型での解析を行ったが, 
%さらに他のSMの拡張模型においても$\mu_\text{min}$や$\beta_2$と模型のparameterの関係をつければ適用可能な解析になっている.

Finally, we comment on the problem of the unitarity~\cite{Burgess:2009ea,Barbon:2009ya,Burgess:2010zq,Hertzberg:2010dc,Giudice:2010ka,Lee:2014spa,Ren:2014sya}.
%最後に, unitarityの問題~\cite{Burgess:2009ea,Barbon:2009ya,Burgess:2010zq,Giudice:2010ka,Ren:2014sya} についてコメントする.
The problem of unitarity does not threaten the consistency of the Higgs inflation by itself.
Concretely speaking, the physical momentum scale during the inflation, which is given by the de Sitter temperature $H_\text{inf}\simeq 10^{14}\GeV\paren{r/0.2}^{1/2}$, is smaller than the \blue{unitarity violation scale} \magenta{$M_P/\xi$ that is evaluated on the electroweak vacuum}.\footnote{\pink{
In Ref.~\cite{Ren:2014sya}, it is shown that %the unitarity violation scale for Higgs inflation is always above $M_P/\sqrt{\xi}$ for $\xi = O(10-100)$. Thus 
the unitarity constraints are relieved for moderate values of non-minimal coupling, $\xi = O(10-100)$.
%the Higgs inflation is also safe from the unitarity bound for 
}}
\blue{
In general, a new physics would appear around the unitarity violation scale.
}
%The cutoff scale in the current electroweak vacuum is given by $M_P/\xi$.  
\blue{It is interesting that
it is around the GUT or string scale in our model. }
%much higher than that in the original Higgs inflation model with $\xi\sim10^4$. 
%In this respect, the Higgs inflation from the SM criticality is \red{an improved scenario}. %compared to the original \red{one with a large $\xi\sim 10^{4}$}.
%original Higgs inflationと同様に, unitarityの問題はinflationのconsistencyを壊さない.
%具体的には, Inflation中の物理的なscaleはde Sitter温度の$H_\text{inf}\simeq 10^{14}\GeV(r/0.2)^{1/2}$であるが, これはinflation中のcutoff scaleよりも十分小さいため, 問題ない.
%現在のElectroweak vacuumでのcutoff scaleは$M_P/\xi$で与えられるが, 我々のsaddle pointを用いたHiggs inflationでは$\xi$がoriginal Higgs inflationに比べてかなり小さく, cutoff scaleはPlanck scale近くになっている.

%この点においてHiggs inflation from SM criticalityはoriginal Higgs inflationよりimproveしている.
%Since non-minimal coupling $\xi$ is $O(10)\sim O(100)$ in our model, the problem of unitarity~\cite{Burgess:2009ea,Barbon:2009ya,Burgess:2010zq,Giudice:2010ka,Ren:2014sya} is significantly improved. 
%%%%%%%%%%%%%%%%%%%%%%%%%%%%%%%%%%%%%%%%
\section*{Acknowledgement}
The work of Y.H.\ is supported by \purple{the} Grant-in-Aid from the Japan Society for the Promotion of Science (JSPS) Fellows No.~25.1107.
S.C.P.\ is supported by Basic Science Research Program through the National Research Foundation of Korea funded by the Ministry of Science, ICT $\&$ Future planning (NRF-2011-0029758) and (NRF-2013R1A1A2064120).
The work of K.O.\ is in part supported by the Grant-in-Aid for Scientific Research Nos.~23104009, 20244028, and 23740192. 

%%%%%%%%%%%%%%%%%%%%%%%%%%%%%%%%%%%%%%%%
%\appendix
%\section{stringから}
%本論文で扱った模型についてstring theoryの立場から考察してみる.
%string frame (全体に$1/g_s^2$がかかり, あとはstring scale $M_s$で次元合わせ)で
%\al{
%S=\int d^4x\sqrt{-g}\frac{M_s^2}{g_s^2}
%\left(
%R\left(1+\xi \frac{\varphi^2}{M_s^2}+...\right)
%+\frac{1}{2M_s^2}\partial_\mu \varphi \partial^\mu \varphi
%+\frac{\lambda}{4}\frac{\varphi^4}{M_s^2}+f \frac{\varphi^6}{M_s^4}
%\right).
%}
%$\xi, \lambda, f$の大きさはもしこれらのcouplingがtree levelで現れるなら$O(1)$が期待され,
%loop inducedならもっと小さくなる(どれくらい？).
%$\varphi$を新たに$g_s \varphi$と再定義することでscalar fieldはcanonicalになり,
%\al{
%S=\int d^4x\sqrt{-g}\left(\frac{M_s^2}{g_s^2}R+\xi \varphi^2 R+
%\frac{1}{2}\partial_\mu \varphi \partial^\mu \varphi
%+\frac{\lambda}{4}g_s^2 \varphi^4+f g_s^4 \frac{\varphi^6}{M_s^2}
%\right).
%}
%が得られる. 
%この論文のprescriptionIの議論で$\xi\sim10, \lambda\sim10^{-6}, f\sim 10^{-8}$の場合にinflationがうまくいくことを見た.このことは, $\xi$はstringのtree levelから, $\lambda, f$はloop inducedで出てきていることを示唆しているように思われる. こうした可能性をより具体的なstringの模型で考えてみるのも面白いだろう.
%\al{
%\int\frac{d^d k}{(2\pi)^d}=\frac{1}{(4\pi)^{d/2}}\frac{2}{d}\frac{1}{\Gamma(d/2)}\Lambda^d
%}

\bibliographystyle{TitleAndArxiv}
\bibliography{Higgs}
\end{document}